\documentstyle[preprint,tighten,aps,epsf]{revtex}

\def\centereps#1#2#3{\vskip#2\relax\centerline{\hbox to#1{\special
  {eps:#3 x=#1, y=#2}\hfil}}}
\def\NI{\noindent}
\def\eqcite#1{(\ref{eq:#1})}
\def\eq#1{Eq.~\eqcite{#1}}
\def\ltsim{\hbox{\kern.25em\raise.5ex\hbox{$<$}\kern-.75em\lower.5ex \hbox{$\sim$}\kern.25em}}
\def\calT{{\cal T}}
\def\ontop#1#2{{#1\atop#2}}
\def\undertext#1{$\underline{\hbox{#1}}$}
\def\xA{x_{\ssc A}} \def\yA{y_{\ssc A}} \def\xB{x_{\ssc B}} \def\yB{y_{\ssc B}}
\def\ssc{\scriptscriptstyle }
\def\lsfig#1#2#3{\epsfxsize=#1in \epsfysize=#2in \epsffile{#3.eps}}

\begin{document}

\begin{titlepage}
{\par \small \hangindent 3 true in \hangafter -5 
\setlength{\baselineskip}{10pt} ~ \newline
For the proceedings of the conference, \'Equations aux D\'eriv\'ees Partielles et Physique Math\'ematique, in honor of J. Vaillant, Paris, June 2000. \par}

\vspace{3.5cm}

\setlength{\baselineskip}{16pt}

\begin{center}
{\LARGE A compromised arrow of time}

\quad
\bigskip

{L. S. Schulman\\
Physics Department, Clarkson University \\
 Potsdam, NY 13699-5820, USA \\
schulman@clarkson.edu \\
\smallskip
\date{\today}}

\vspace*{.5cm}

{\bf Abstract}\\ \end{center}

\setlength{\baselineskip}{12pt}  

{The second law of thermodynamics---the usual statement of the arrow of time---has been called the most fundamental law of physics. It is thus difficult to conceive that a single dynamical system could contain subsystems, in significant mutual contact, possessing opposite thermodynamic arrows of time. By examining cosmological justification for the usual arrow it is found that a consistent way to establish such justification is by giving symmetric boundary conditions at two (cosmologically remote) times and seeking directional behavior in between. Once this has been demonstrated, it is seen that entropy increase can be reversed and that the usual arrow is less totalitarian than previously believed. In the same vein, other boundary conditions, modeling shorter periods in the evolution of the cosmos, can be found that allow the simultaneous existence of two thermodynamic arrows, notwithstanding moderate interaction between the systems possessing those arrows. Physical consequences of the existence and detection of opposite-arrow regions are also considered. 
}
\end{titlepage}

\setlength{\baselineskip}{14pt}

\section{Introduction}

The thermodynamic arrow of time is so all-encom\-passing that, paradoxically, it can be invisible. The ancients did not explain friction, just as they did not think to try to explain why things fell {\it down\/}. There were myths for spring and myths for rain, but as far as I know, no myths for terrestrial gravity or friction. Nevertheless, once Newton had formulated reversible laws of mechanics (by studying the nearly frictionless motion of the planets) it was realized that the enormous asymmetries of our lives---breaking eggs, graying beards, heating brake pads---required an explanation that went beyond the laws of dynamics.

It is not always appreciated that this explanation involves two separate logical steps. The first is to go from reversible {\it microscopic\/} dynamics to irreversible {\it macroscopic\/} manifestations. A result of this kind is the Boltzmann H-theorem. The second step takes the first as given, but then asks, why do we have the particular arrow that we do, what is it that picks the actual direction? Is it a physical effect, like CP violation or the expansion of the universe? Or perhaps it's more subtle. Any sufficiently large system must have {\it some\/} arrow, so that the directionality question loses its meaning. One would still need to establish, however, that there must be but a single arrow. Background on both aspects can be found in \cite{timebook}, although emphasis in the present article and in \cite{opposite,paradox,reply}
is on the second matter.

The study of irreversibility is closely related to ergodic theory and the tools in the present study are the tools of that field. In particular I use the model systems popular there, one of the most congenial of these being the ``cat map." I will first use this to show how an arrow of time can {\it apparently\/} develop in a symmetric system and how this can be related to cosmology. Then a system with two arrows will be exhibited. Next I will treat the important physical question of whether we could hope to see this phenomenon. Finally I will take up one of the amusing aspects of this research, the possibility of causal paradoxes, very much like those that arise in time travel scenarios.

\section{The cat map}

The ``cat map" is a mixing transformation of the unit square onto itself. As for ``Schr\"odinger's cat," the whimsical name was acquired when a fundamental idea was illustrated by having unpleasant things happen to a cat. For $x,y\in I^2=$ the unit square, the mapping $\phi:I^2\to I^2$ is 
\begin{eqnarray}
      x'=\phi_1(x,y)\equiv &\ x+\phantom{2}y \hbox{~~mod~1}&  \nonumber\\
      \noalign{\vskip -30pt} \nonumber\\                    % was 28
              &  &  \qquad\qquad   
				  \hbox{or~~~}
      \pmatrix{x'\cr y'\cr}\equiv
            \pmatrix{1&1\cr 1&2\cr}\pmatrix{x\cr y\cr} \hbox{~~mod~1} \label{eq:catmap}\\
      \noalign{\vskip -25pt} \nonumber\\
      y'=\phi_2(x,y)\equiv &\ x+2y \hbox{~~mod~1}&  \nonumber\\\nonumber
\end{eqnarray}
The matrix on the right has determinant one, so that $\phi$ is measure preserving. If one thinks of $I^2$ as phase space, then this is a model of classical mechanics. Technically, $\phi$ is effective at producing apparently irreversible effects because the matrix in \eq{catmap} has as its larger eigenvalue $\lambda\equiv \left(3+\sqrt{5}\right)/2 \sim 2.6>1$.

But it is the non-technical illustration of the relaxation that earned the map its name. In a classic text \cite{arnold}, an initial pattern consisting of an image of a cat is shown. When $\phi$ is applied once, the image is stretched by a factor $\lambda$, squeezed transversely by $1/\lambda$, snipped (when overshooting $I^2$), and reassembled. Within two or three applications of $\phi$, the poor cat is hardly recognizable.

Feline images will not be needed in this article. Rather I place several hundred points in $I^2$ and allow each to move independently under $\phi$. This models a collection of non-interacting atoms, an ``ideal gas," coming to equilibrium---or failing to do so, as the case may be.

To gain a quantitative handle on relaxation and to define an ``entropy," it is necessary to introduce ``coarse grains" in $I^2$. Some such device is always necessary (see \cite{timebook}), since if one knows precise coordinates and microscopic dynamics there is no loss of information, hence no entropy increase. The coarse grains are here obtained by placing a grid over $I^2$ and taking as the {\it macroscopic\/} state {\it only\/} the number of points within each rectangle defined by the grid. For $N$ ($\gg1$) identical atoms in $G$ coarse grains, the entropy is
\begin{equation}
S = -\sum_{k=1}^G \rho_k \log \rho_k \qquad\hbox{with}\quad \rho_k=n_k/N
\label{eq:entropy}
\end{equation}
where $n_k$ is the number of points in coarse grain (rectangle) \#$k$.

In the next section the time dependence of the entropy will be illustrated. If all points are in a single rectangle, then $S=0$; when the points are Poisson distributed $S \ltsim \log G$. The relaxation time to go from a single $0.1\!\times\!0.1$ grain to equilibrium is about 5. Using smaller grains, one can see that $S(t)\sim t\log\lambda$ (and $\log \lambda$ is the Lyapunov exponent).

\section{Cosmology and two-time boundary value problems}

Around 1960 Gold \cite{gold} argued that the thermodynamic arrow of time followed the expansion of the universe. In other words, your coffee cools because the quasar 3C273 recedes. Lest this reformulation test your credibility, let me give perspective. The expansion of the universe is another factual arrow of time, a ``cosmological" arrow. Lacking any great asymmetry in the {\it laws\/} of nature, it is plausible that the two all-pervasive arrows, cosmological and thermodynamic, could have something to do with each other. To make the case you need to show how the distant and unnoticed-until-1930 cosmological arrow could induce the thermodynamic one. Gold argued that the expansion of the universe provides a sink for free energy, by accepting---due to its expansion---all radiation thrown off by stars, etc. He also showed how the influence of large scale processes reached to smaller scales, even to coffee at Caf\'e Epsilon. He clinched his argument with a {\it gedanken\/} experiment in which a star was put in an insulated box and held there till it equilibrated. At this point the contents of this large box no longer had a thermodynamic arrow. He then imagined that he would open a window in the box for a short time, during which radiation would {\it escape\/} to the universe at large, this escape (rather than its opposite) being the consequence of expansion (cf.\ Olber's paradox). With the box again closed and with some of its contents gone, the system within is no longer in equilibrium. The subsequent return to equilibrium is exactly what provides the sought-for arrow and the connection to the larger cosmological scenario. A consequence of Gold's ideas is that in a contracting universe (which occurs for some solutions of the equations of general relativity) the arrow would be reversed.

Back in 1972 I was totally enamored of this argument. But now one learns why it is useful for physicists to talk to mathematicians. I tried to explain Gold's argument to a mathematician colleague of mine, Andrew Lenard, and mid-story came to a halt. Trying to maintain his high standards of logic I realized that Gold's star-in-a-box argument was circular.

First a counterexample: suppose the radiation had {\it entered\/} the box (as in a contracting universe, a l\'a Olber). According to the tale above, there would {\it still\/} be an arrow in the box after the window closed, as the system recovered from the influx of radiation. It would need to recover from {\it any\/} change. So something is clearly fishy.

Now if Gold's arrow in his star-in-a-box parable does {\it not\/} arise from the efflux/influx distinction, where does it come from? The answer is, it is the arrow of the {\it narrator}. In listening to this story one automatically accepts that prior to the opening of the window there is no effect of the opening. {\it But that is already an assumption about the arrow of time!\/} If this same process of opening and closing had been observed by someone with an opposite arrow (a notion one must allow if one is engaged in establishing the existence of one or the other arrow) that observer would naturally have assumed that the effect of that process would have occurred on the other side of the interval within which the window was open, namely an epoch we (or the narrator) would consider to be {\it before\/} the opening. 

How can this circularity be avoided? I proposed \cite{correlating} that one should give boundary conditions at two remote times and look for the emergence of an arrow between them. For example, if our universe has both a big bang and a big crunch (expansion followed by contraction), then the following sort of boundary condition would fit this prescription. Take the matter and radiation distributions as roughly homogeneous at some time interval $\tau$ both after the big bang and before the big crunch. Such homogeneity appears to have obtained at the epoch of radiation-matter decoupling, with $\tau\sim$ 300,000 years. Now include the expansion and contraction and try to derive that at first (after time-$\tau$) there would be an arrow in one direction, then---as the big crunch becomes nearer in time than the big bang---the other. This does {\it not\/} mean that we must live in such a universe. It is only a framework within which Gold's idea can be logically explored.

As stated, the project is daunting. But I have taken the usual route of statistical mechanicians, namely to abstract the problem, retaining the essential conceptual features while making the mathematics tractable.

The model system is a ``gas" of particles each of which has cat map dynamics in the phase space $I^2$. How to model the expansion of the universe? The essential feature (for our purposes) of a rapid expansion from a homogeneous configuration is that homogeneity is a far, far from equilibrium state when the dynamics of the system are dominated by gravity. Gravity makes things clumpier; it {\it enhances\/} density differences. So the homogeneous matter and radiation distribution that prevailed at $\tau$---and which we will demand prevail at $\calT - \tau$ (with $\calT$ the big crunch time)---is far from equilibrium, or it will become that as the universe expands \cite{kepler}.

For the cat map gas, the way to be out of equilibrium is to {\it not\/} be uniformly distributed, and for a given coarse graining this can be accomplished by putting all the points in a single grain. As observed earlier (following \eq{entropy}) this gives the system entropy $S=0$. But now we want {\it two}-time boundary conditions, and we want them symmetric. So we pick a pair of times, 0 and $T$, and demand the same sort of nonequilibrium state for the gas at both these times.

Solving two-time boundary value problems can be difficult. Imagine a similar problem for the gas in a 40~m$^3$ room. Demand that all the gas be in a single cubic meter at time-0 (so the rest of the room is in vacuum), and require that under particle-particle interactions alone (and with the room totally isolated) all the gas spontaneously find itself in that same cubic meter one hour later. It is extremely unlikely that the exact initial data for the gas would lead to this eventuality, although {\it some} initial conditions do reach the desired time-$T$ state. Finding them is another matter. If you randomly chose initial velocities and positions within the required cubic meter, a conservative estimate suggests that you would have less than one chance in $10^{10^{26}}$ to have all the particles arrive where you want them an hour later. For the ideal gas though the job is easy. I want a gas of 250 ``atoms" and a coarse graining that divides $I^2$ into 50 rectangles. The trick is to randomly place 12,500 points in the desired initial rectangle and let these atoms evolve $T=19$ time steps. The result is a square fairly uniformly covered with points. Then, {\it discard all points that do not fall into the desired final rectangle.\/} (For an ideal gas this does not affect the dynamics of those that remain.) This leaves you with about 250 successful points (1/50th of 12,500).

% FIGURE 1
\def\szx{1.833}
\def\szy{1.5}
\def\sx{-.6 true in}

\vbox{
\vskip 5 pt
\hbox to 6 true in {\vrule  width 6 true in height .4pt depth 0pt}
\hbox{   \hskip -.15truein
\lsfig{\szx}{\szy}{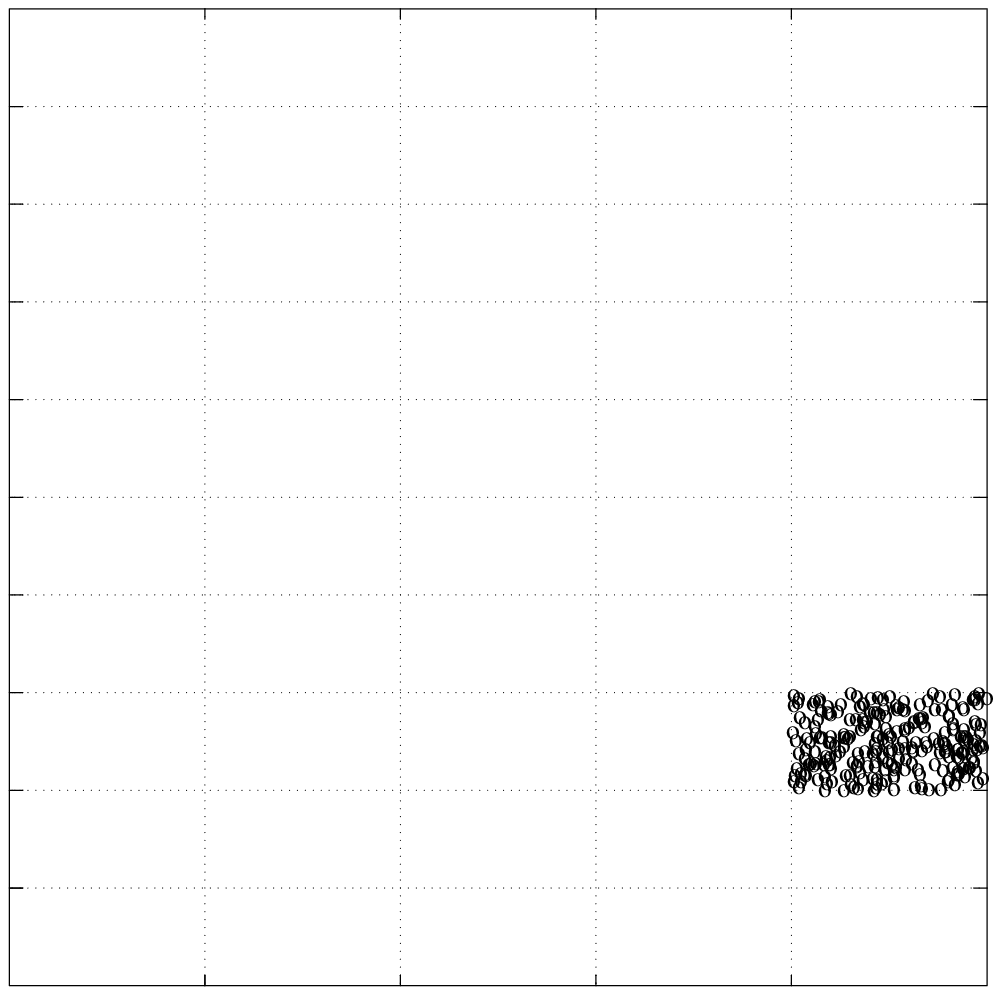}\hskip \sx
\lsfig{\szx}{\szy}{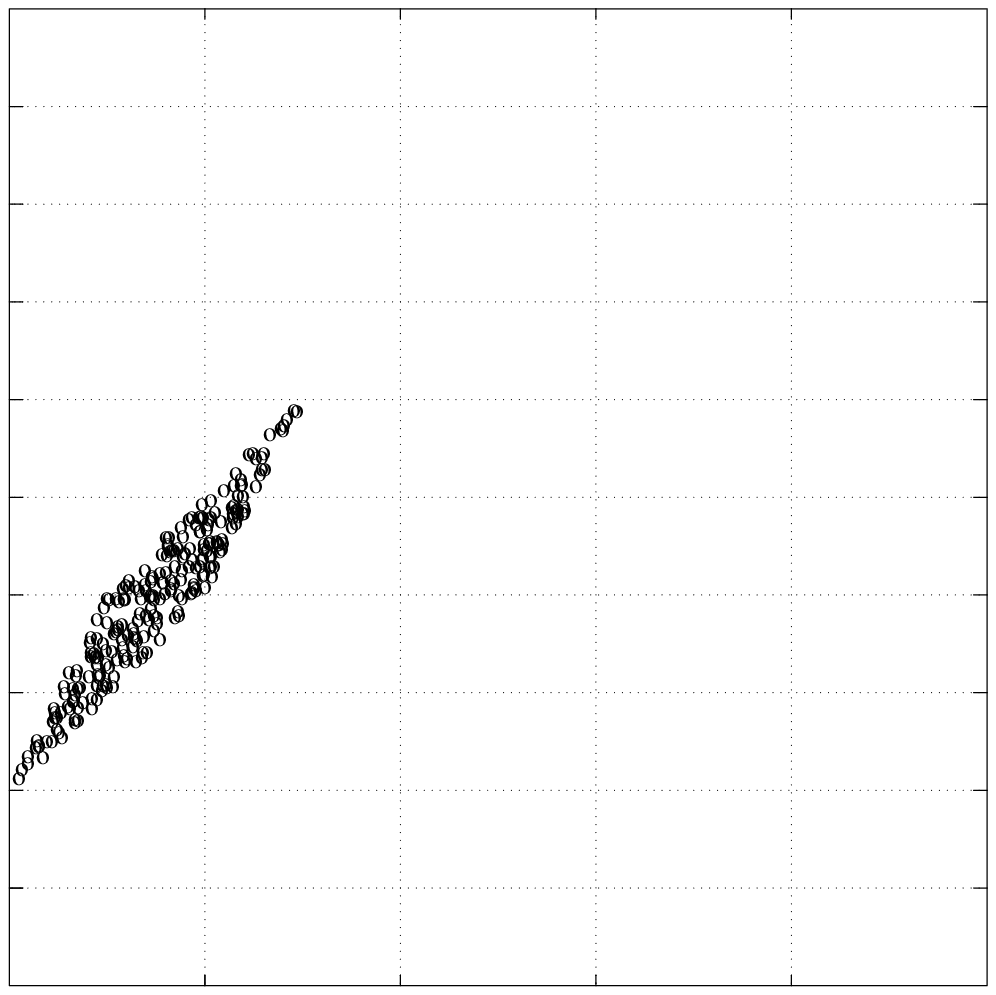}\hskip \sx
\lsfig{\szx}{\szy}{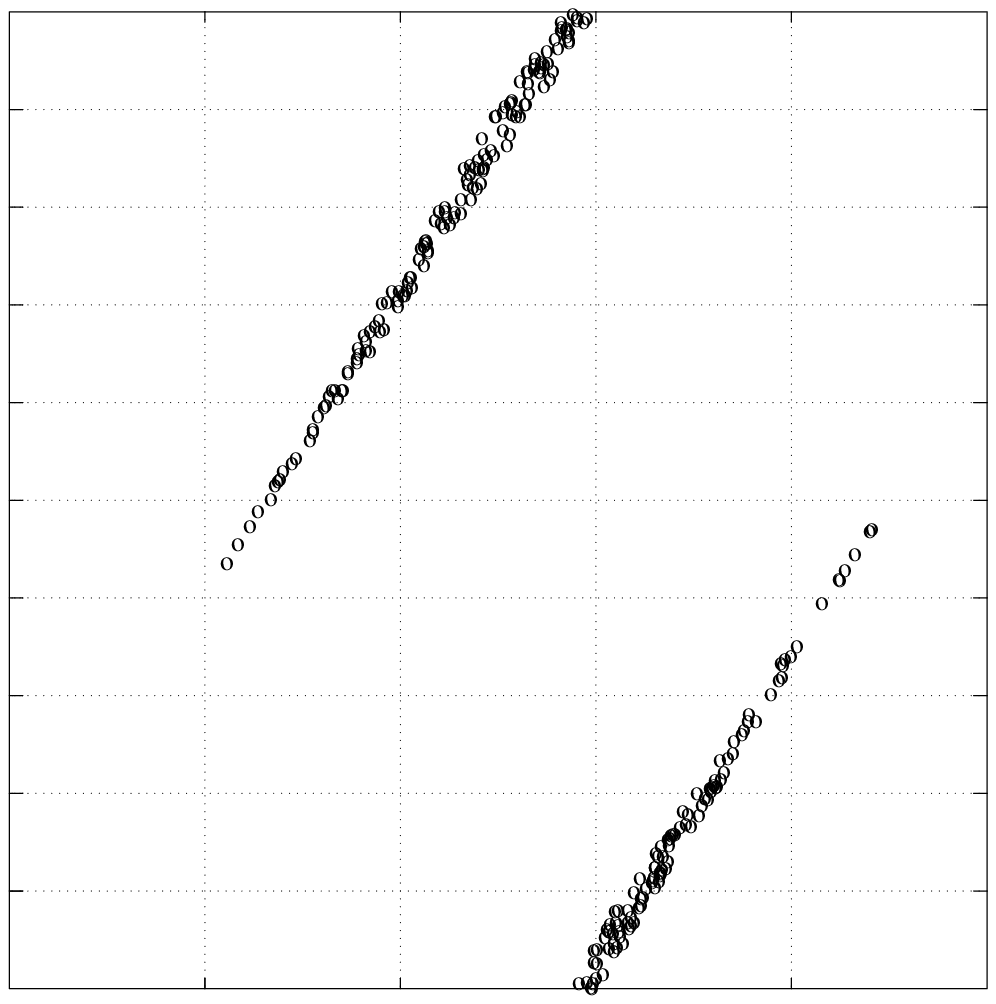}\hskip \sx
\lsfig{\szx}{\szy}{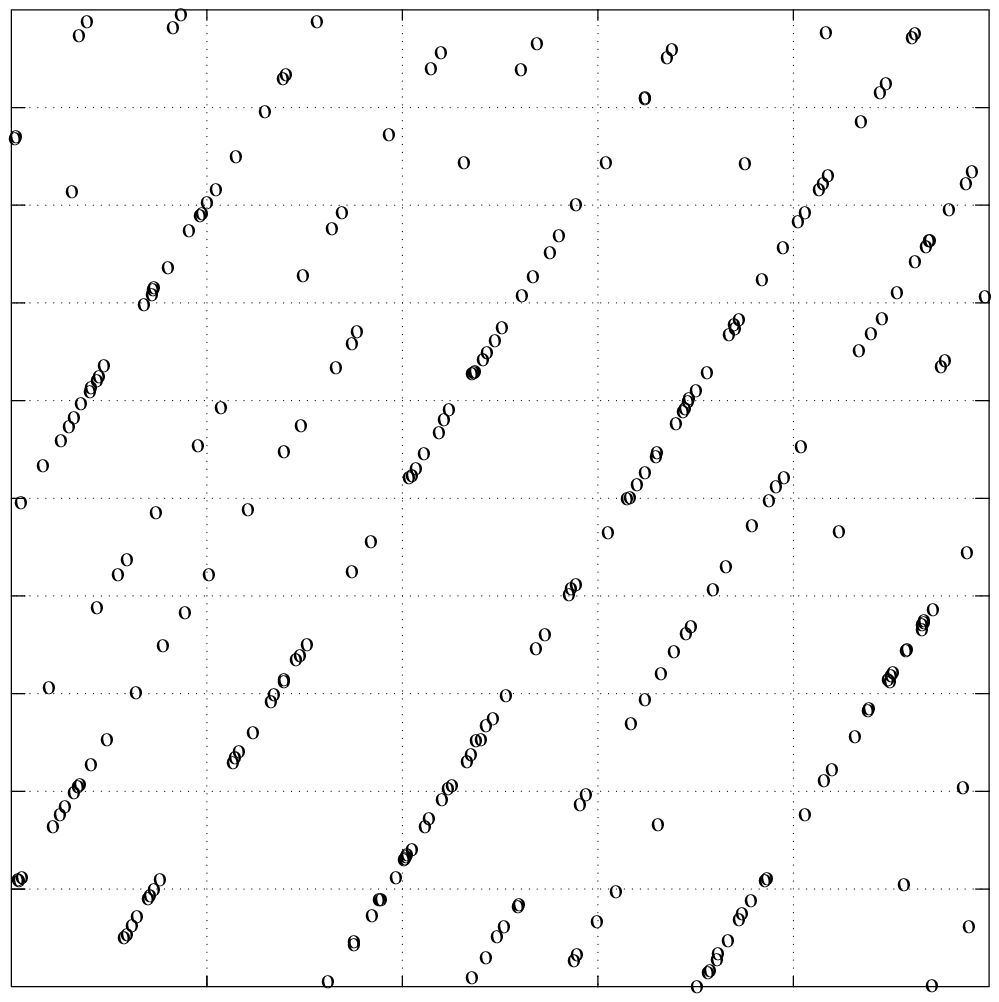}
}

\hbox{\hskip -.1truein
\lsfig{\szx}{\szy}{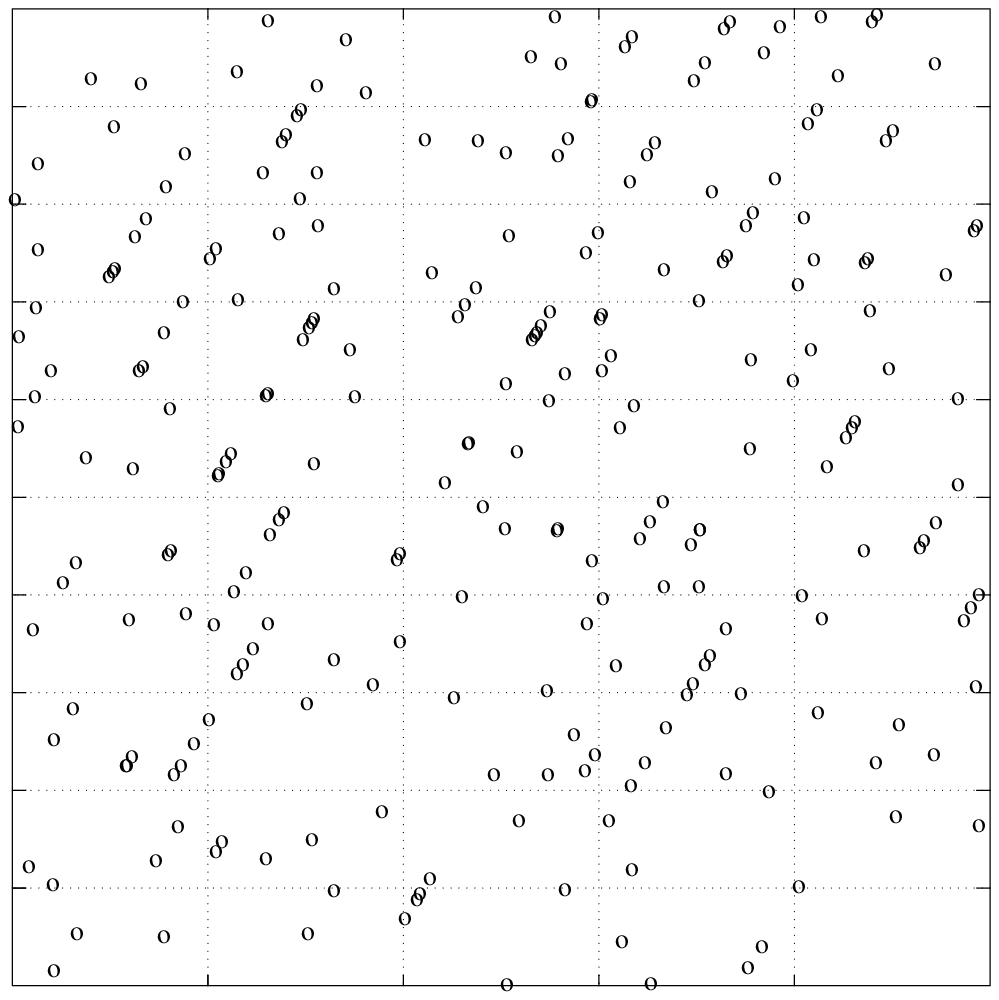}\hskip \sx
\lsfig{\szx}{\szy}{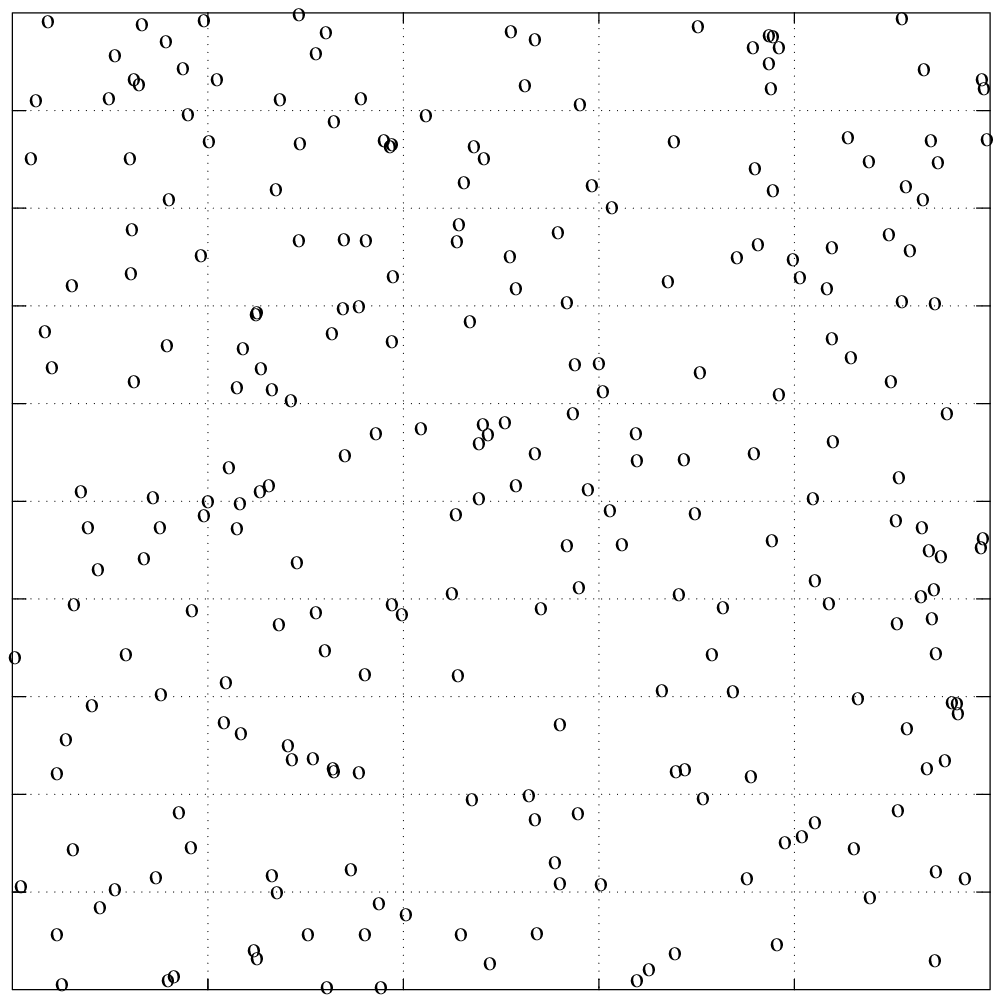}\hskip \sx
\lsfig{\szx}{\szy}{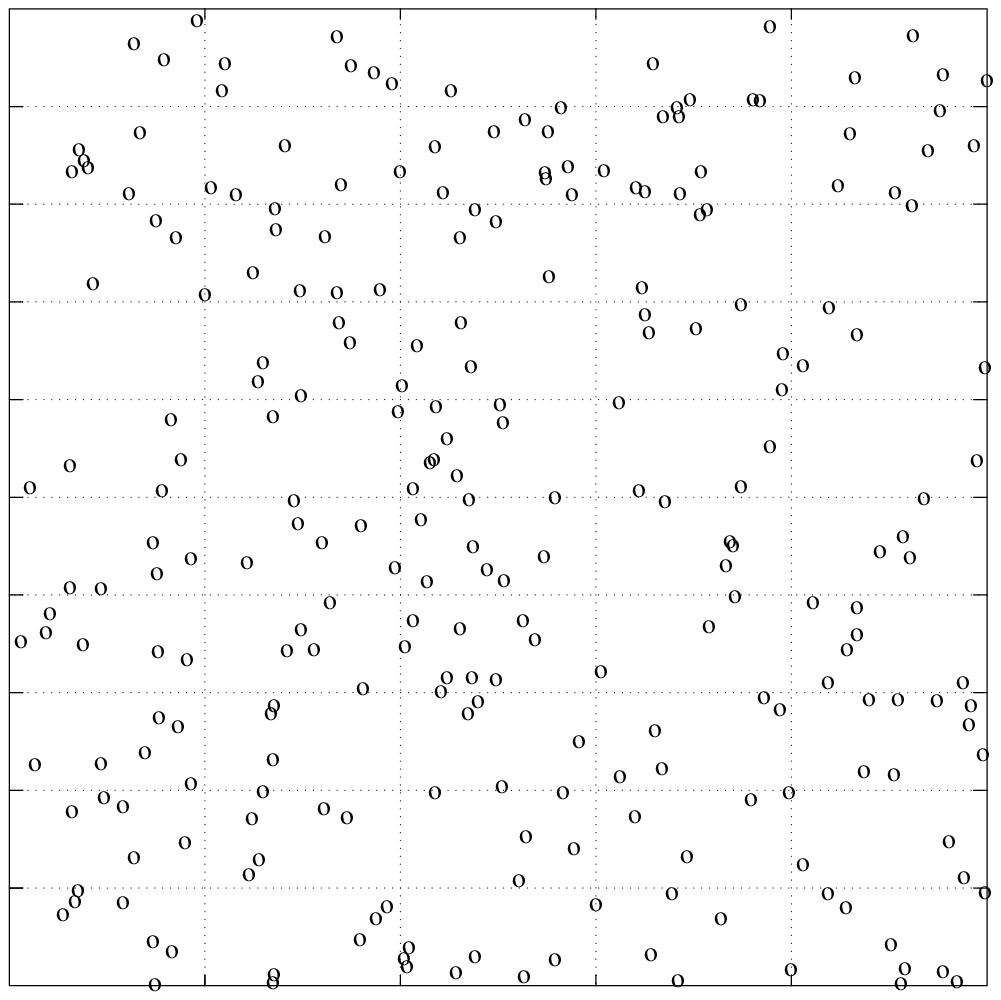}\hskip \sx
\lsfig{\szx}{\szy}{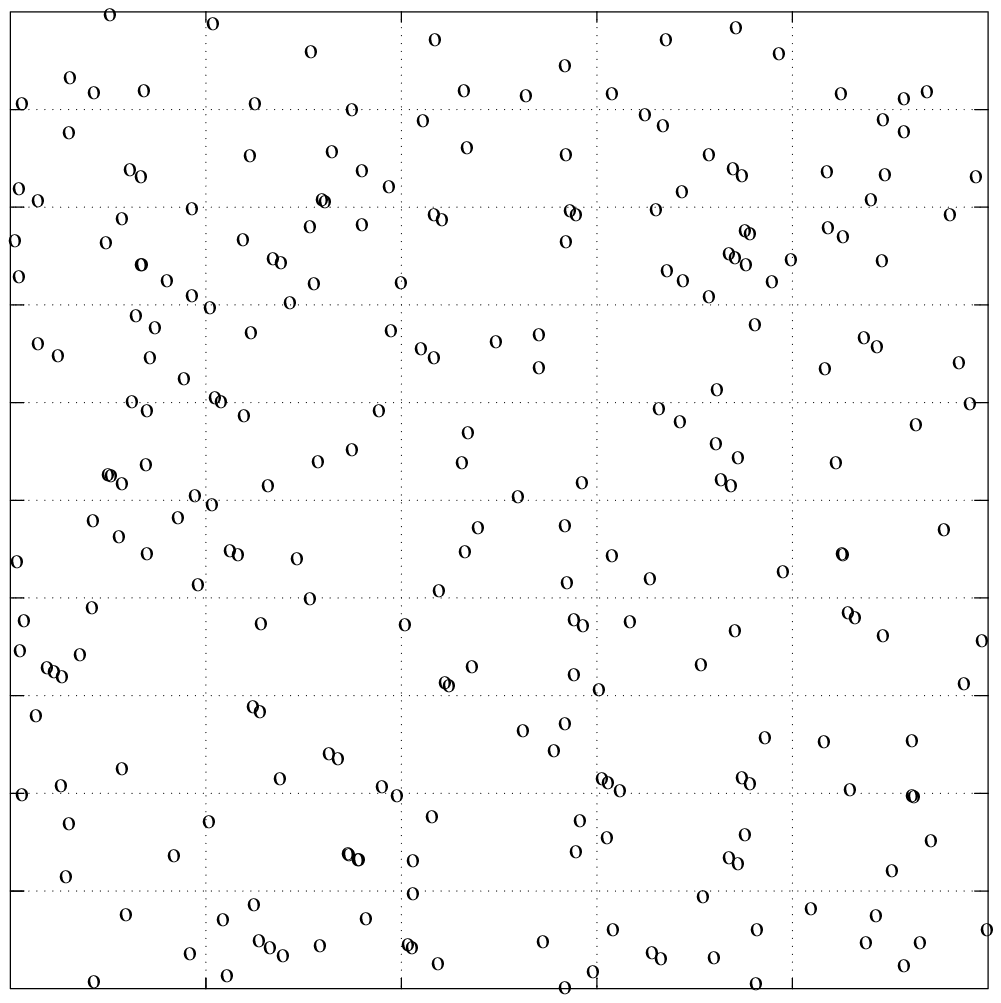}
}

\hbox{\hskip -.1truein
\lsfig{\szx}{\szy}{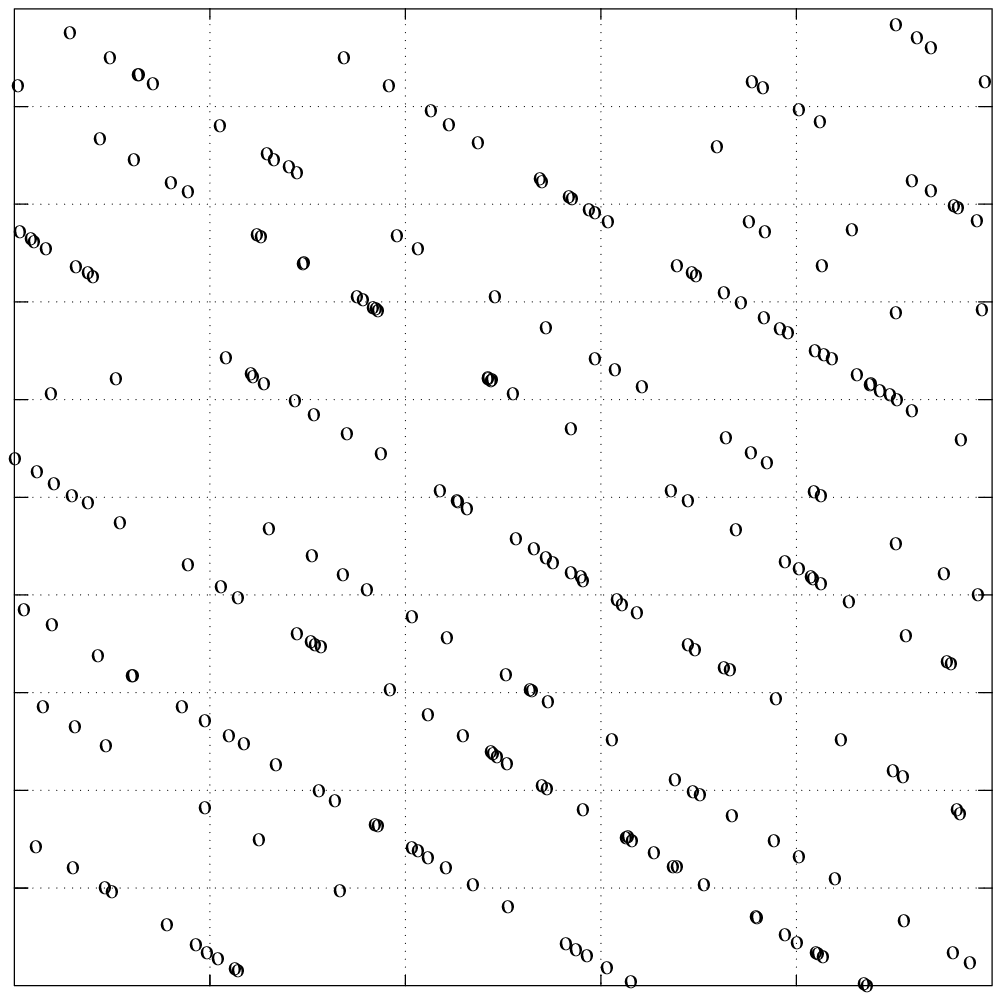}\hskip \sx
\lsfig{\szx}{\szy}{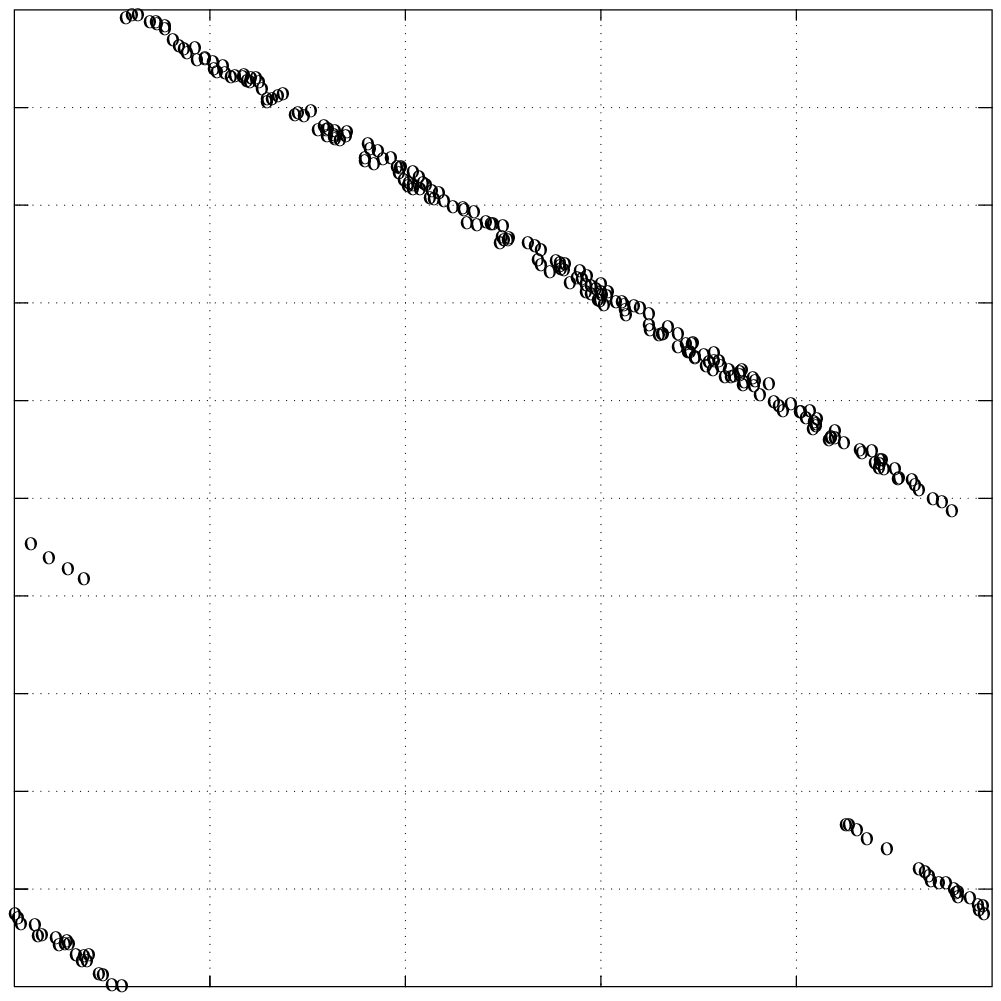}\hskip \sx
\lsfig{\szx}{\szy}{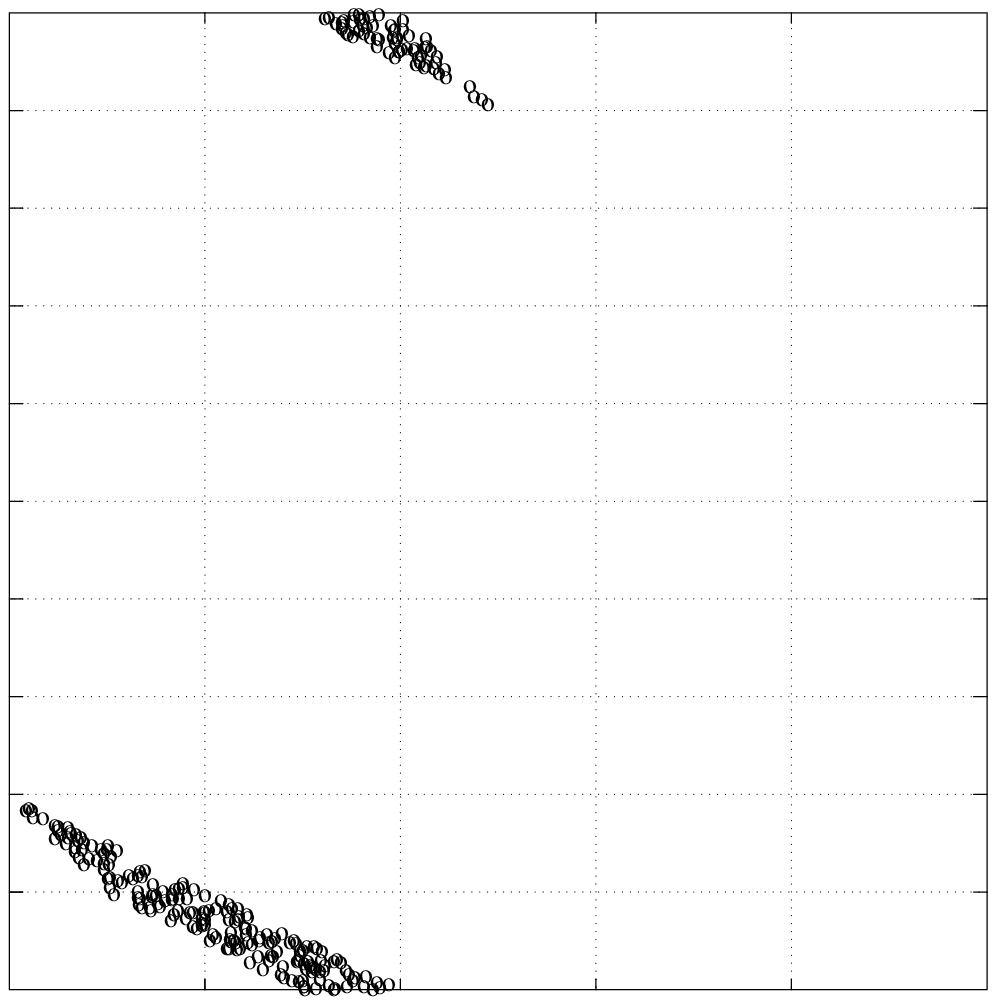}\hskip \sx
\lsfig{\szx}{\szy}{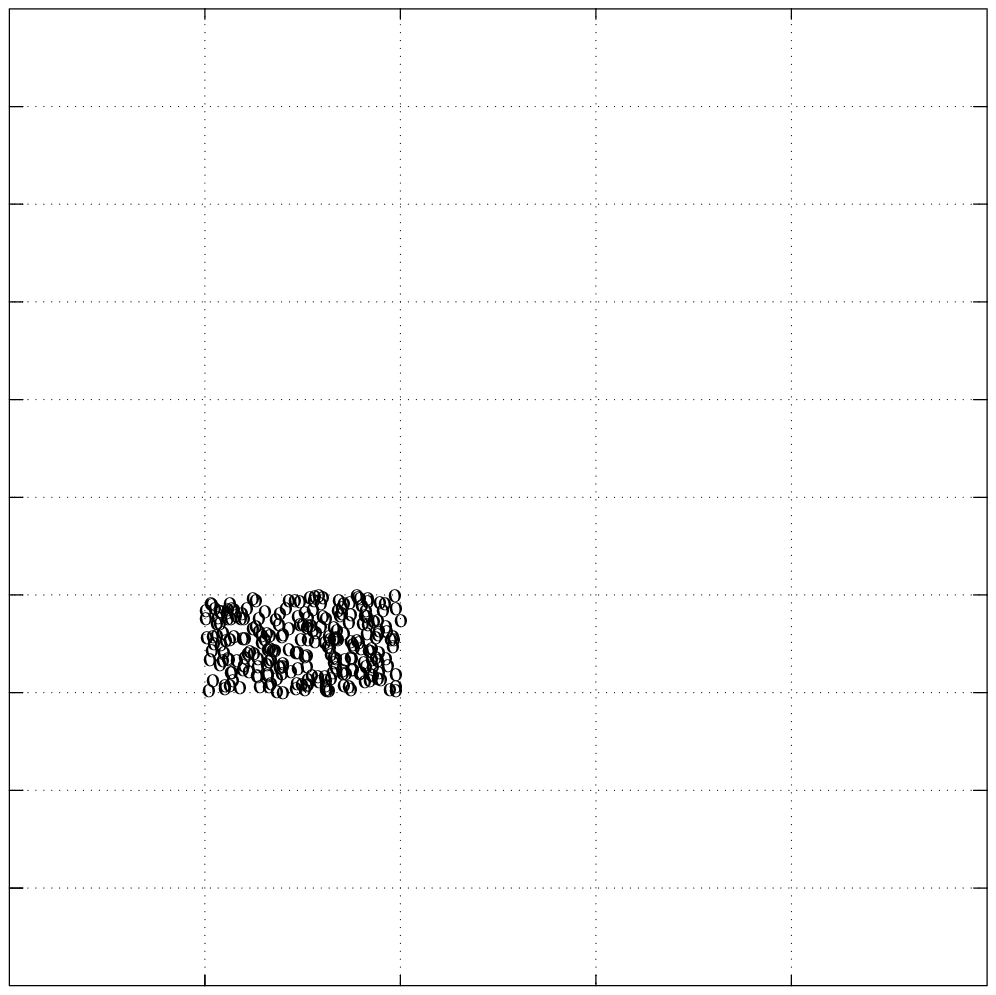}
}
{\NI Figure 1. History of 250 points that begin in a single coarse grain at time-0 and end in a single coarse grain at time-19. Selected times are shown
according to the following scheme: top row (left to right), \hbox{[0 1 2 4]};
second row, \hbox{[5 8 11 14]}; third row, \hbox{[15 17 18 19]}.}
\vskip 1 pt
\hbox to 6 true in {\vrule  width 6 true in height .4pt depth 0pt}
%\label{movie}
%\end{figure}
}
% End FIGURE 1

In Fig.~1 is a ``movie" of the history of these points, with 12 selected time steps shown. Note the first half of the movie: if I would show you the expansion of a gas with {\it no\/} future boundary condition you would not be able to tell the difference. To illustrate this, in Fig.~2a I show the entropy as a function of time for a gas of 250 particles under free expansion, and in Fig.~2b is the time dependence of the entropy for the actual ``movie." Compare the initial increases in entropy for the two situations: they are indistinguishable, up to statistical fluctuations. For this {\it macroscopic\/} quantity, the patterns of entropy increase are essentially identical. A second point is that if Fig.~2b were flipped about its middle ($t=9.5$) it would look the same (up to statistical fluctuations).

% FIGURE 2
\vbox{
\vskip 5 pt
\hbox to 6 true in {\vrule  width 6 true in height .4pt depth 0pt}
\def\sz{3}
\def\sx{-.1 true in}
\hbox{                       % \hskip -.1truein
\lsfig{\sz}{\sz}{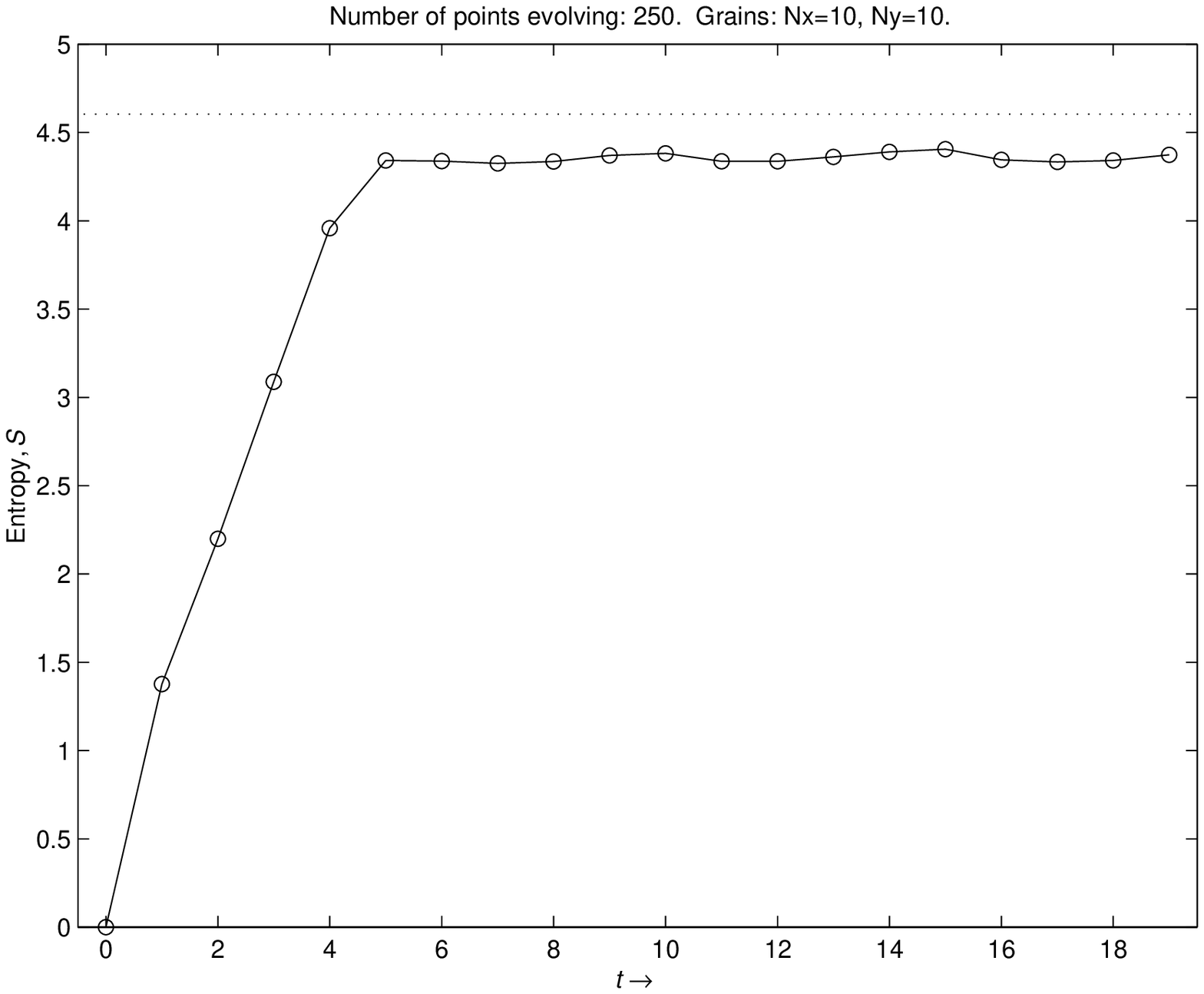}   % \hskip \sx
\lsfig{\sz}{\sz}{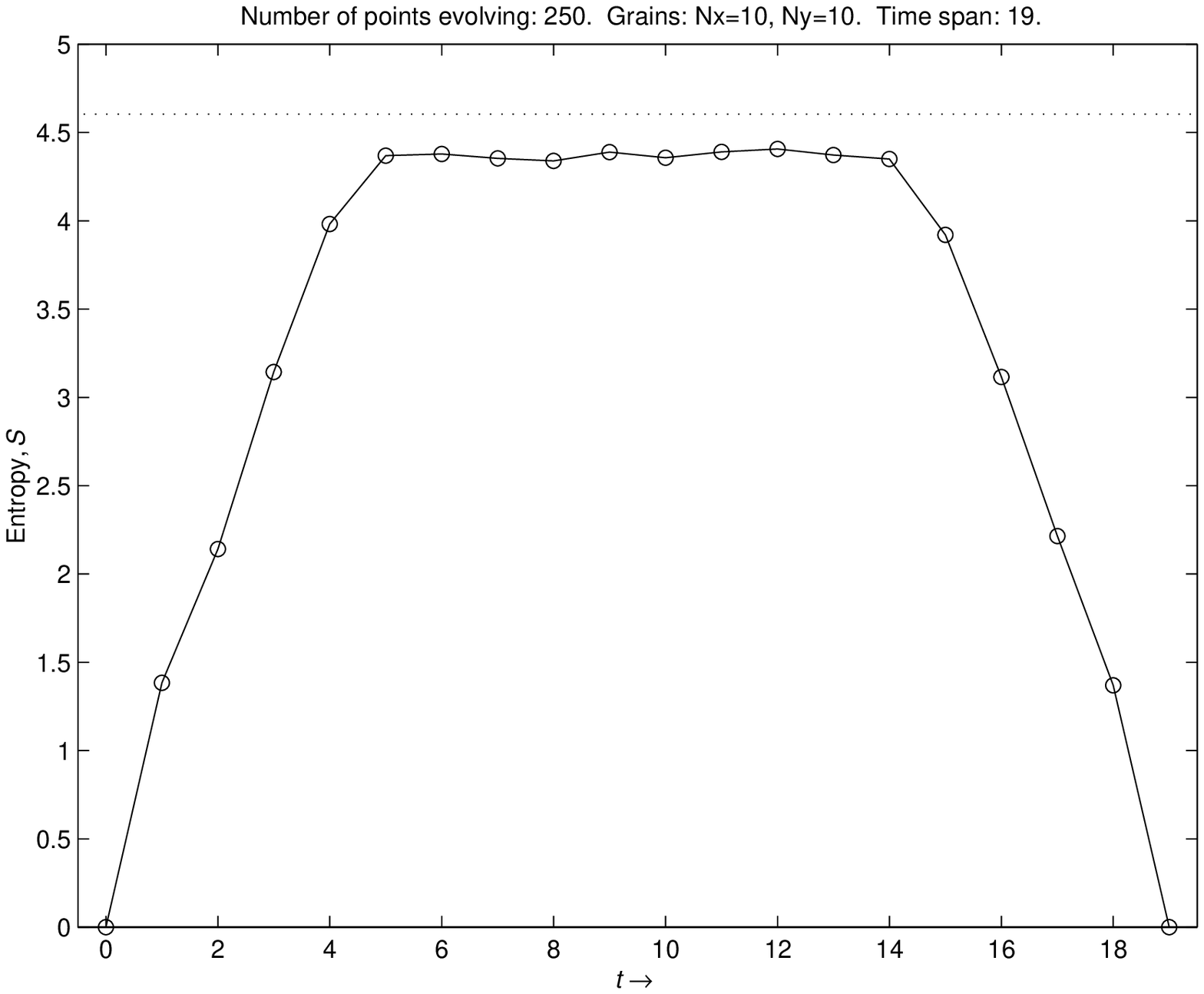}
}
\NI Figure 2. Entropy, $S$, as a function of time. The left figure (2a) is $S(t)$ {\it without\/} future conditioning. On the right (2b) is $S(t)$ for a two-time boundary condition solution, such as Fig.~1. Note the similarity of the initial time dependence. Also of interest is the reflection symmetry of Fig.~2b about its temporal midpoint.
\vskip 1 pt
\hbox to 6 true in {\vrule  width 6 true in height .4pt depth 0pt}
}
% End FIGURE 2

Now let's apply the message of this simulation to the grand issue of cosmology. Both boundary times, 0 and 19, are states of low entropy. This is analogous to the state of the universe long enough after ``decoupling" (or before, coming back from the big crunch) for the homogeneous distribution no longer to represent equilibrium. For early times (in the cat movie) the entropy increases and indeed there is an arrow of time. As remarked, this arrow is indistinguishable from what you would have if there were \undertext{\it no\/} future conditioning. Thus we have no way of knowing from this statistical information alone whether there is or is not a big crunch coming. The point of this demonstration is that the expansion of the universe creates a situation in which the homogeneous distribution, which is disordered---and \undertext{likely}---during one epoch, becomes extremely \undertext{{\it un}likely} at a later epoch (once gravity dominates). As the larger system proceeds to its more disordered state, as it {\it relaxes,\/} it is temporarily trapped in all sorts of metastable states, for example stars, which in turn relax and drive shorter term metastable states, such as people. Note too, from the symmetry about the midpoint in Fig.~2b, that the arrow toward the end is reversed.

\section{4. Opposite arrows}

So far we have talked about ``the" thermodynamic arrow of time. Could it happen that within some large system there are {\it two} subsystems, each with a different arrow of time? To see how this might come about imagine that we do indeed live in a big bang-big crunch universe. Now get a well-insulated spaceship and put yourself in a cryogenic bath for a long time---until the universe has begun to contract. Now your alarm clock awakens you and you look out the porthole to see the world around you with an opposite-running arrow! Similarly I can imagine that there could be matter in regions that have been relatively isolated ``since" the big crunch and which ``survive" \cite{quotesense} into our epoch.

But could there really be two arrows at once, entropy increasing in one region, decreasing in another? Here is an argument against this possibility. From the standpoint of one of the observers (a sentient being in one component) the events in the other region are entropy {\it lowering.\/} An example of this would be if someone carefully arranged the positions and velocities of 11 billiard balls so they would come together, with 10 of them forming a triangle of balls at rest and the eleventh going away from them along one of the axes of the triangle (the reverse of a ``break"). This would be a system with decreasing entropy; it becomes {\it more} ordered. Now if such a system is disturbed even slightly, someone coughing while the balls collided, it would ruin the enormous coordination required for the entropy lowering process. So for an observer in either region to see (and maybe yell in surprise at) the creation of order in the other, would be to destroy that order. It follows that opposite running arrows cannot coexist if systems interact.

The foregoing argument is wrong. It is given from the perspective of one observer, and implicitly assumes that the way to formulate a macroscopic problem is as an {\it initial\/} value problem in the time sense of that observer. It is true that for an initial value problem even small perturbations will disturb entropy lowering, but we should use arguments that do not already assume the validity of one arrow. As discussed, the way to do this is to give macroscopic boundary values at {\it two} times, and solve for the motion between.

This leads to the following formulation for producing simultaneous opposite arrows of time. Consider a system with two component subsystems; call them A and B. Give macroscopic boundary conditions such that A's entropy at time-0 is lower than it is at time-$T$, and B's entropy has the reverse specification. A and B {\it are\/} allowed to interact 
\cite{slice}.

What sort of solution can be expected in the time interval $[0,T]$? If there were {\it zero} interaction, this would be equivalent to two separate problems. A's entropy would increase as a function of the time parameter $t$, while B's would decrease. This is a trivial consequence of the boundary conditions and the only anomaly is that B made a poor choice of time parameter. This emphasizes that ``$t$" is a parameter for microscopic time evolution, with no a priori relation to thermodynamics.

Now allow interaction. Do the arrows survive? I investigated this with an extension of the cat map. Take two squares, A and B (copies of $I^2$), and allow $N$ particles in each to evolve under the cat map. In addition, each point in A or B is associated with a particular point in the other box, and is influenced by the position of that other point. To describe the map efficiently introduce additional notation. Recall that the original cat map (\eq{catmap}) is called $\phi$. Now define a map, $\psi_\alpha(u,v) \equiv(u+\alpha v,v)\;\hbox{mod}\,1$, depending on the real parameter $\alpha$. Like $\phi$, $\psi$ is measure preserving, but the eigenvalues of the corresponding matrix are both unity, so it is not chaotic. For each time step, the combined motion of each pair of pairs $[(\xA,\yA),(\xB,\yB)]$ is a three-step process: 1) A-B interaction: $\psi_{\alpha/2}$ applied to $(\xA,\yB)$ and $(\xB,\yA)$ separately; 2) the usual cat map evolution: $\phi$ applied to $(\xA,\yA)$ and $(\xB,\yB)$ separately; 3) repeat step \#1 \cite{practice}.

\def\startfig{
\vskip 5 pt
\hbox to 6 true in {\vrule  width 6 true in height .4pt depth 0pt}
}

\vbox{
\startfig
\centerline{$\hbox{\large A, ~~\it t~}\longrightarrow$}
\def\sz{1.1} \def\sx{-.3 true in}
\hbox{\hskip -.1truein
\lsfig{\sz}{\sz}{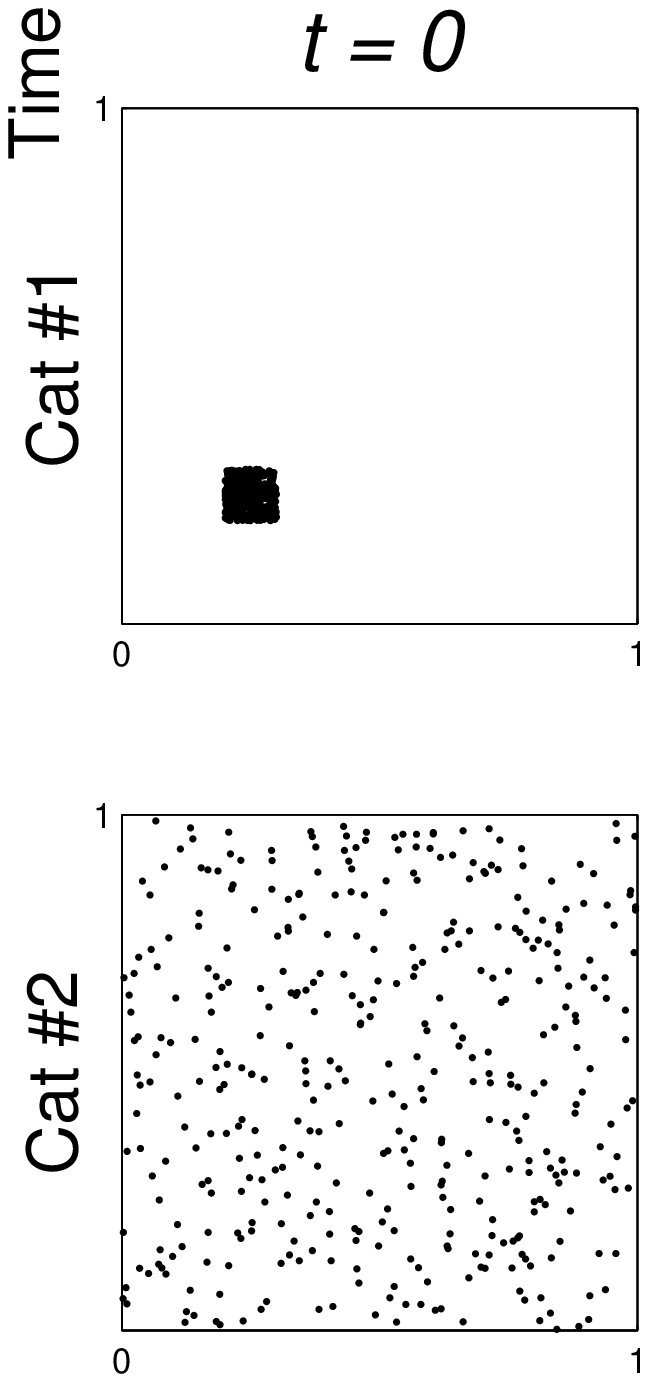}\hskip \sx
\lsfig{\sz}{\sz}{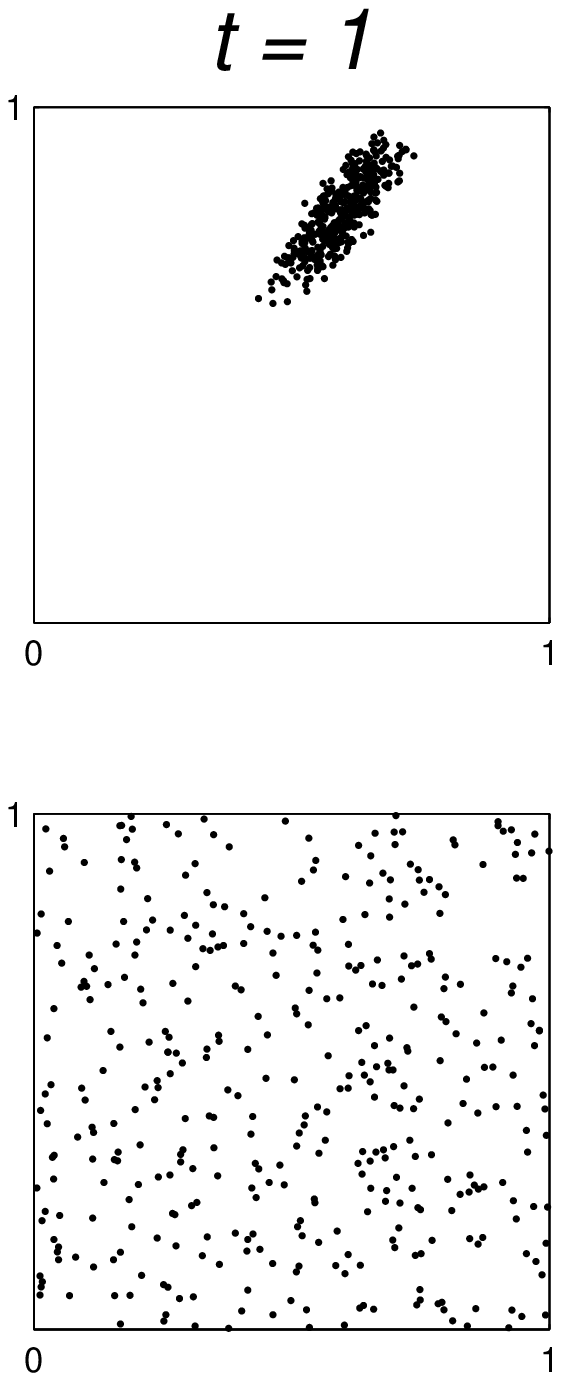}\hskip \sx
\lsfig{\sz}{\sz}{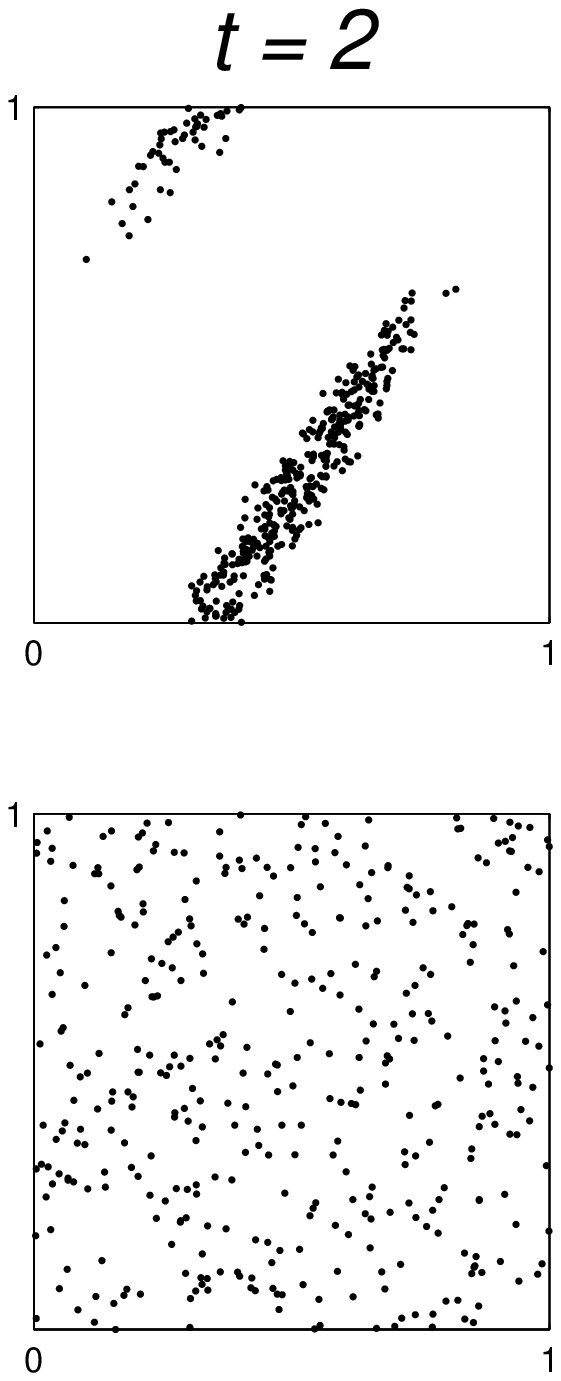}\hskip \sx
\lsfig{\sz}{\sz}{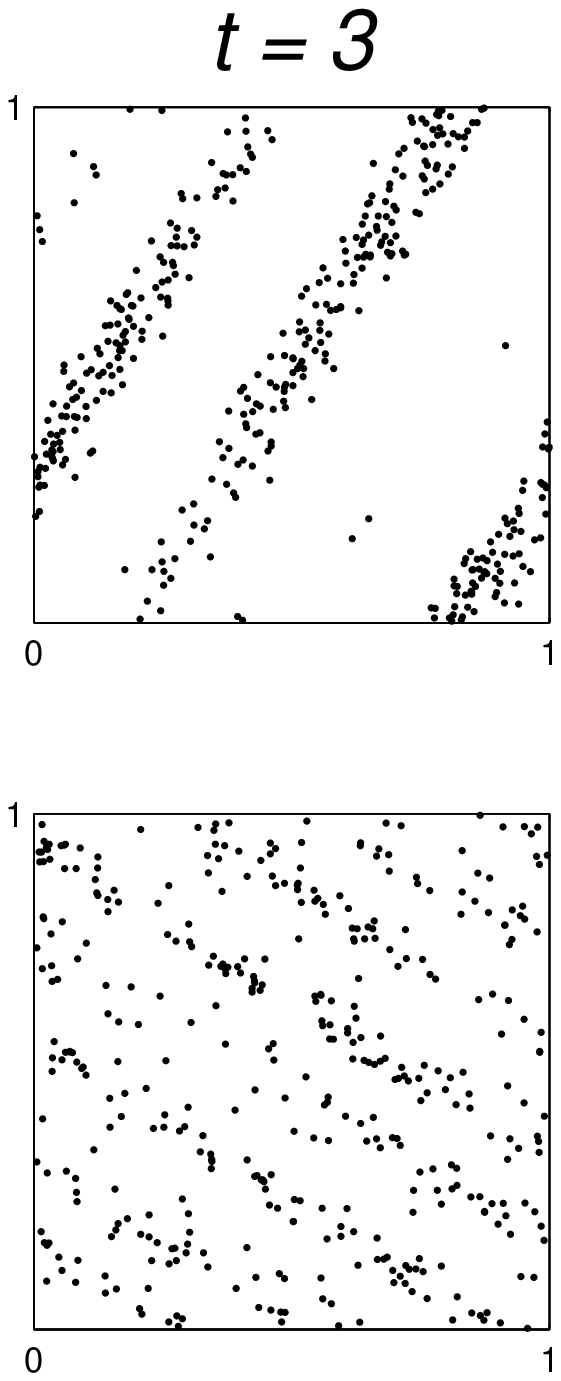}\hskip \sx
\lsfig{\sz}{\sz}{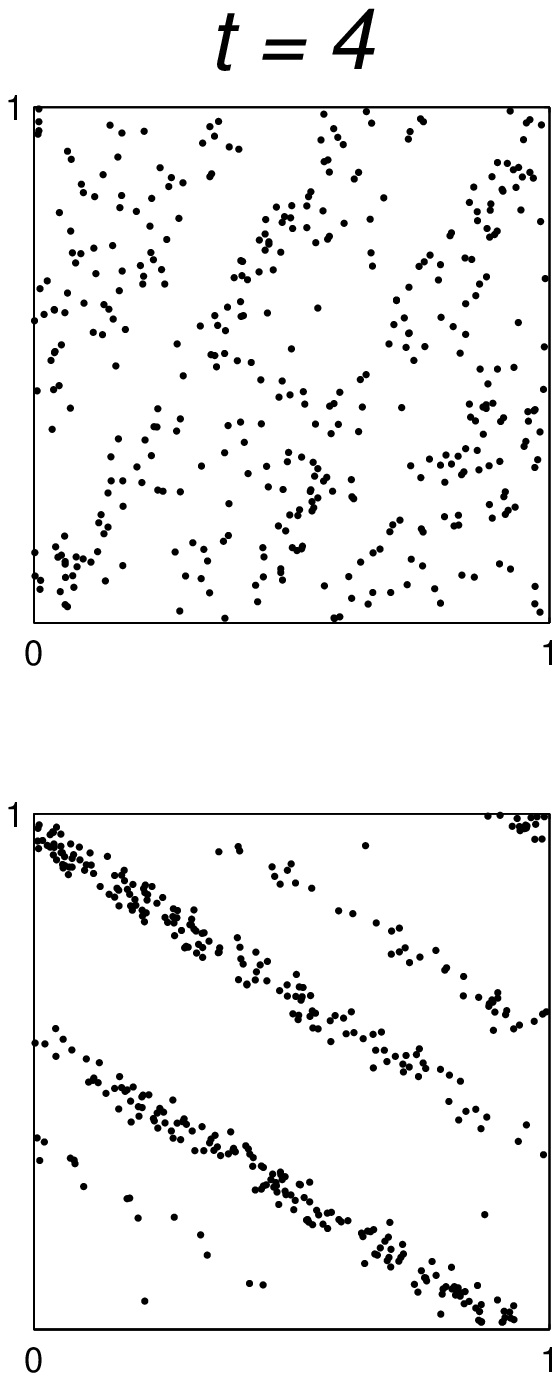}\hskip \sx
\lsfig{\sz}{\sz}{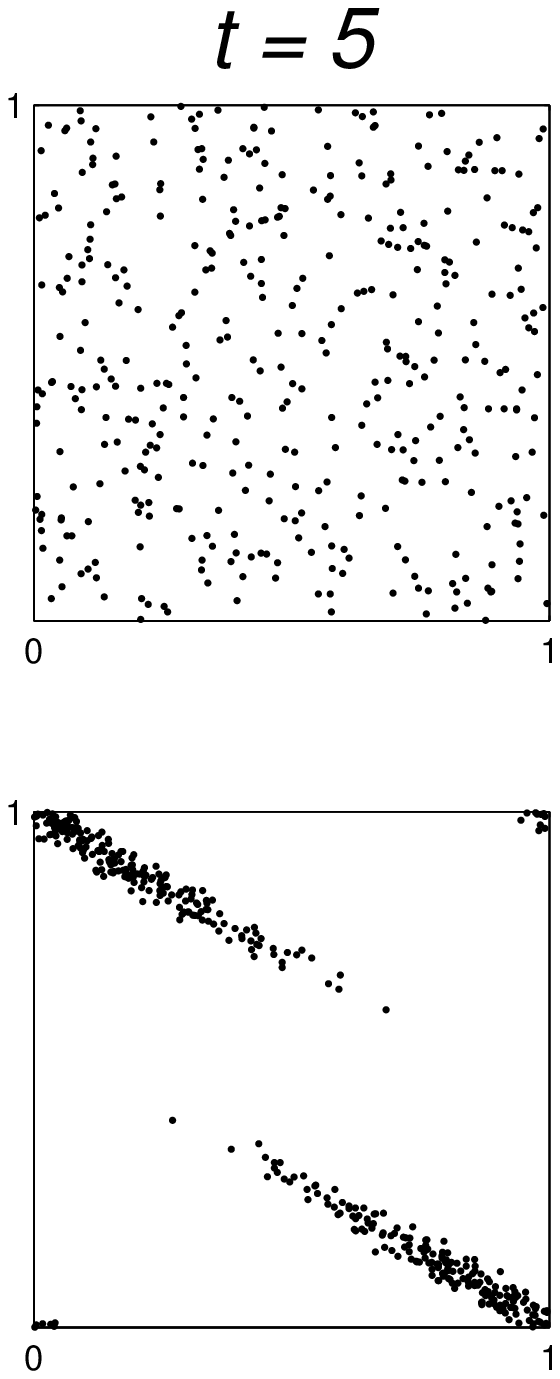}\hskip \sx
\lsfig{\sz}{\sz}{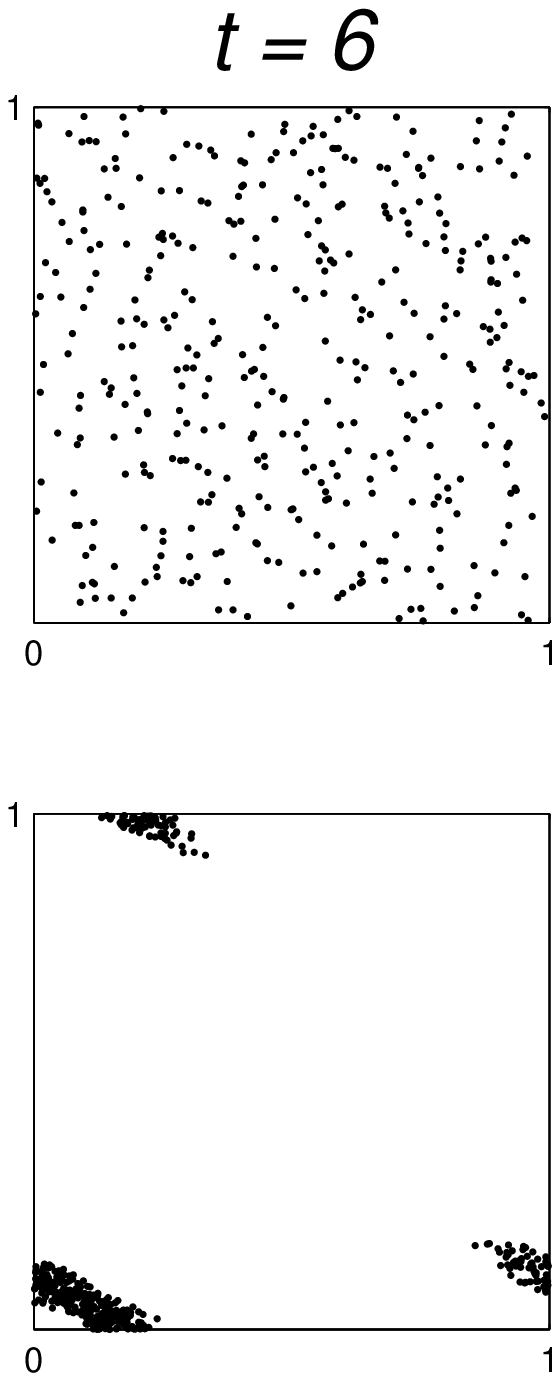}\hskip \sx
\lsfig{\sz}{\sz}{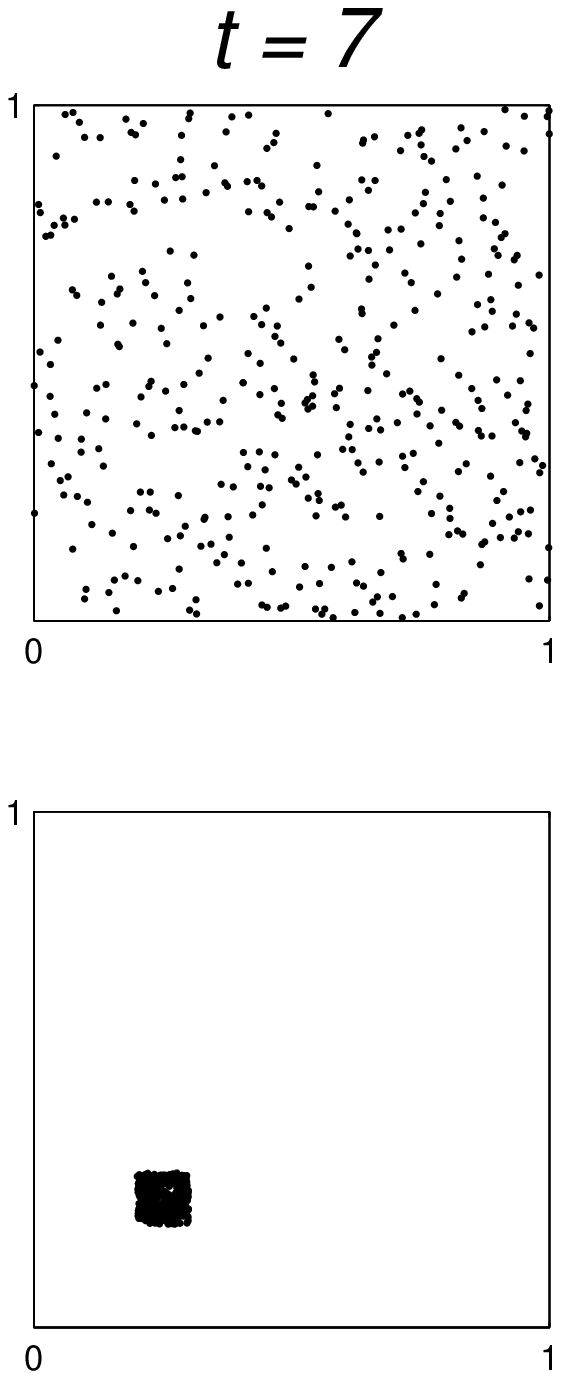}
}
\vskip .1 truein
\centerline{$\longleftarrow \hbox{~\large\it t},~~~ \hbox{\large B}$}
\smallskip
\NI Figure 3. Images of a pair of interacting systems, each having a different arrow of time. For the upper row (system A) the boundary conditions are that all points are in a single coarse grain at time zero, with no condition set for time $T$ ($=7$). For the lower row (system B) all points are in a single coarse grain at time-$T$, with no requirement for time-0. The systems are coupled to each other with the parameter $\alpha$ (in the map $\psi_\alpha$) taking the value 0.2.
\vskip 1 pt
\hbox to 6 true in {\vrule  width 6 true in height .4pt depth 0pt}
}

The results of the simulation are shown in Fig.~3. Clearly the the goal of opposing arrows has been realized. It is interesting though to note the effects of the coupling. Without coupling the result of a single time step on the initial rectangle for system A would be a neat parallelogram, the image of that rectangle under the cat map (cf.\ Fig.~1, time step 1). In Fig.~3 it is clear that the parallelogram has been smudged. This is the effect of $\psi$ and system~B. Nevertheless, the overall expansion shows an arrow ($S(t)$ will be displayed momentarily). The effect of B on A is thus seen to be essentially noise, and correspondingly for the effect of A on B.

In Fig.~4L (left) is the time dependence of the entropy (but not for the same run as Fig.~3). Three values of $\alpha$ (in $\psi_\alpha$) are shown. The first is $\alpha=0$, for which as discussed, it is trivial that there should be two arrows. For $\alpha=0.2$ increase and decrease of entropy in each system is quite similar, although looking carefully one can note that the relaxation is slightly quicker. This is consistent with the images of Fig.~3 and with the idea that the effect of each system on the other is noise. Finally the lowest image shows the behavior for the stronger coupling, $\alpha=0.5$. Now the relaxation is rapid, for both systems, showing that with strong coupling the boundary values can force low entropy momentarily, but it is quickly lost.

For Fig.~4R (right) I take up another question: m\undertext{a}croscopic causality. I take this to mean that macroscopic effect follows macroscopic cause, and distinguish it from m\undertext{i}croscopic causality, stated, e.g., in terms of field commutators. Defining a noncircular test of (macro) causality requires caution. If we were to use initial conditions, then the effect of a perturbation would {\it by definition} be subsequent.

\vbox{
\startfig
\def\tempsize{2.9}
  \hbox to 4.5 true in{          \vsize=2 true in
        \vbox to 3 true in {   \hsize=3 true in
          \hskip -.35 true in
          \lsfig{\tempsize}{\tempsize}{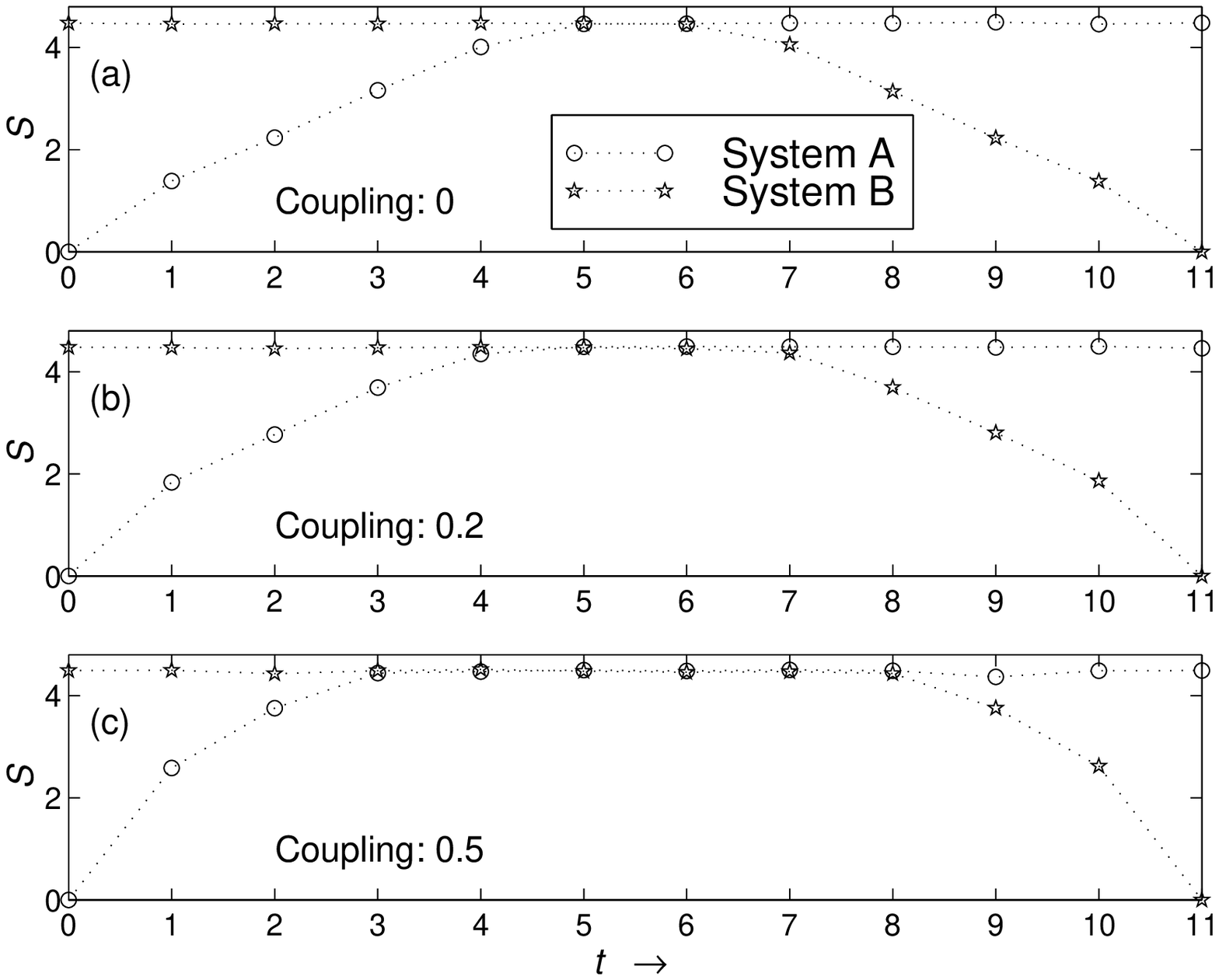} 
                     }
        \vbox to 3 true in {   \hsize=3 true in
           \hskip -.3 truein   % was 2.5
           \lsfig{\tempsize}{\tempsize}{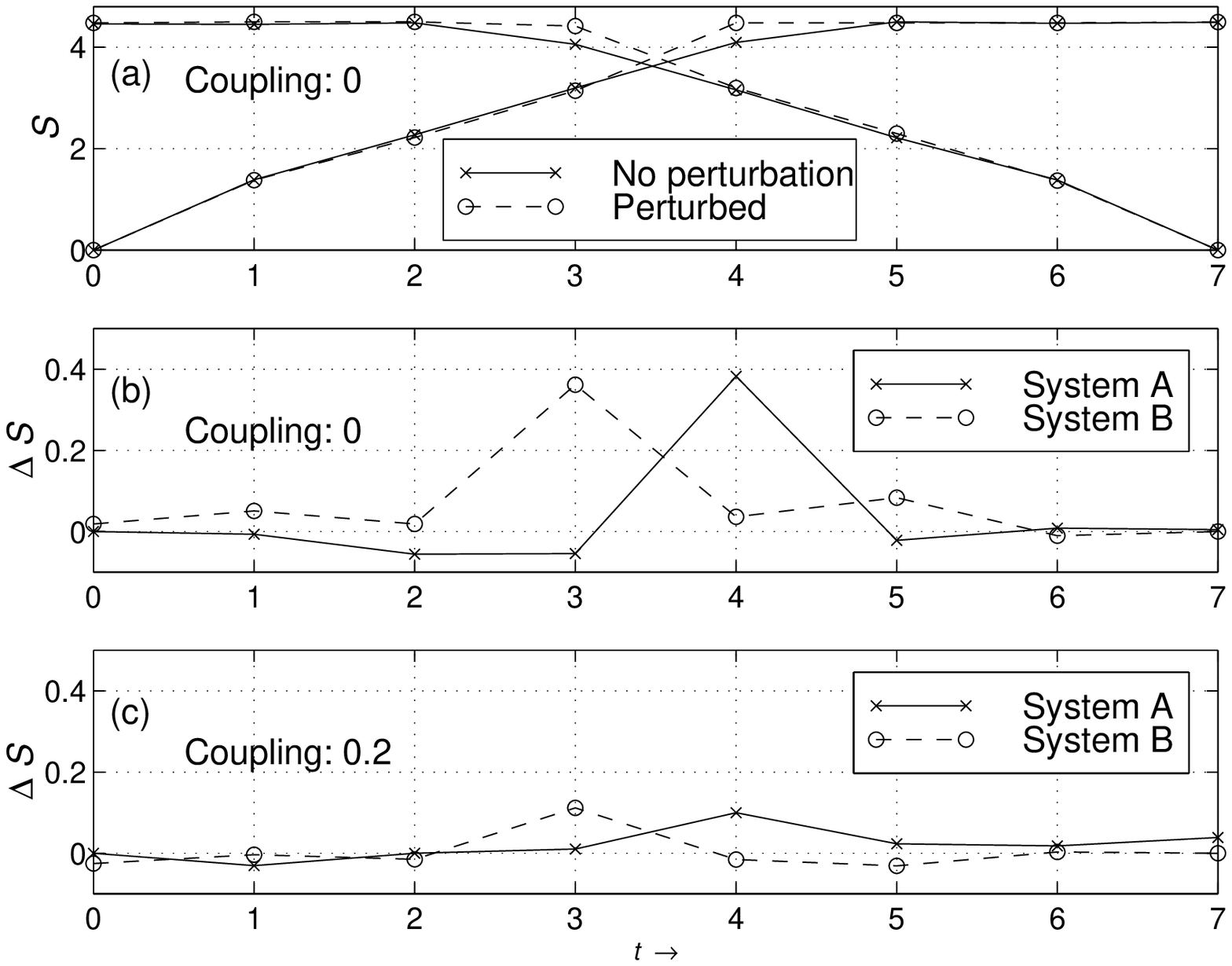}   
                     }
                        }          % close original hbox
\NI Figure 4. Entropy, $S$, as a function of time. The left figure (L) depicts $S(t)$ for three values of the parameter $\alpha$. The uppermost (a), $\alpha=0$ shows the non-interacting system and gives, as it trivially should, increasing entropy for A (as a function of the non-thermodynamic time parameter $t$), and decreasing entropy for \hbox{B}. Below that (b \& c) are shown $S(t)$ for $\alpha=0.2$ and 0.5, respectively. On the right (R) are shown entropy and entropy differences when the system is perturbed. The perturbation takes place effectively at time 3.5 and for A the changes in macroscopic behavior (whether for $\alpha=0$ or 0.2) occur {\it later\/} by A's clock. For B the changes take place at lower $t$ values, which are also later---by B's clock.
\vskip 1 pt
\hbox to 6 true in {\vrule  width 6 true in height .4pt depth 0pt}
}

In \cite{timebook} appears a consistent test of causality in the framework of macroscopic two-time boundary conditions (our usual way of maintaining logical consistency in arrow-of-time questions). The system is required to be in particular coarse grains at times 0 and $T$, and is evolved microscopically from initial to final grains with a particular evolution law, yielding some macroscopic history. Next the system is evolved a second time, but now {\it with a perturbation}. The same boundary data are used for both runs. A ``perturbation" consists of changing the evolution law on a single time step, some $t_0$, with $0<t_0<T$. In general the set of microscopic points satisfying the macroscopic boundary conditions without the perturbation will be different from those satisfying the (same) boundary conditions {\it with\/} the perturbation. So except for times 0 and $T$ the points need not occupy the same coarse grains. Now we can state the test of macroscopic causality: if the {\it macroscopic} behavior is the same {\it before\/} the perturbation but not after, there is macroscopic causality. If it is the other way around, there is reverse causality. If the macroscopic behavior is different {\it both\/} before and after, there is no causality. Now apply this test to our elaborated cat map. The perturbation is that on a particular time step ($t_0$), instead of applying $\phi$ and $\psi$, another rule is used. That rule is that at $t_0$, instead of $\phi$, we use a ``faster" cat. That is, the matrix of the usual cat map \eq{catmap},
$\left(\ontop11\ontop12\right)$, is replaced by 
$\left(\ontop32\ontop43\right)$, which has a larger Lyapunov exponent.

In Fig.\ 4Ra an entropic history is shown for uncoupled systems, with, again, $T=7$. The perturbation is nominally at $t_0=4$, but because entropy ($S$) values are calculated between steps, it is best to think of this as $t_0=3.5$. To better see the effect of the perturbation, in Fig.\ 4Rb I show only the entropy {\it change} due to the perturbation. For A the major difference occurs at 4, while for B it is at 3, consistent with causality. As indicated earlier, for uncoupled systems this result is trivial and only confirms our method. In Fig.\ 4Rc, coupling (0.2) is turned on and the same comparison made. Qualitatively causality persists, although the coupling (by inducing noise) reduces all deviations.

This demonstration of causality responds to another question the reader may have had. My evidence for arrows of time in the earlier discussion consisted solely of increase or decrease of entropy. This now shows that such increase or decrease corresponds to deep-rooted concepts of cause and effect.

\section{Physical implications}

The foregoing development shows that it is {\it possible} for there to be two opposite running arrows. But, {\it does it actually happen}? Could we look across the Milky Way and watch coffee unstir, stars unexplode, Humpty Dumpty come together?

This is a two-part question. First, is there opposite-arrow matter out there? Second, if it's there, could you observe it? Norbert Wiener thought you could not \cite{wiener}. Thus I will devote a short subsection to checking this.

\subsection{Communication between opposite-arrow regions}

Using the same modelling methods that established the possibility of contrary arrows, I will now show that a signal can pass from one system to the other. In \cite{opposite} I considered electromagnetic radiation, and used the time-symmetric Wheeler-Feynman absorber theory \cite{absorber} as a framework for studying electromagnetic communication between the regions. That demonstration has subtleties, particularly in connection with the origins of radiation reaction, that I now prefer to avoid. You need not resolve all the old issues of classical radiation in order to see that opposite arrow regions can have a macroscopic influence on one another.

Again we consider a two-time boundary value problem, and again A will be in a single grain at time-0 and B in a single grain at time-$T$. But in this experiment, at a particular intermediate time, $t_0$, I will wiggle \hbox{A}. By ``wiggle" I mean that instead of the usual composite cat map step at $t_0$, the points of A will be shifted. That's all. On that same step, nothing happens to B. To be precise, in the example I worked out for Fig.~5, on time step~4, the points of B were mapped by the identity ($2\times2$) matrix, while those of A underwent the following transformation: $x'=x+0.3$, $y'=y+0.5$, both mod~1. Of course this affects the points selected (to satisfy the two-time boundary value problem) at the microscopic level. But what is the macroscopic effect? To see this, I compare this run to one in which {\it neither} A {\it nor} B does anything on time step~4. (For all these the same boundary conditions were used.) This was done with more points than the earlier work in order to reduce statistical fluctuations. The entropic history of both runs is recorded, and Fig.~5 shows the {\it difference} between the entropies in the two cases.

\vbox{
\startfig
\hbox{\hskip 1 truein
\lsfig{4}{3}{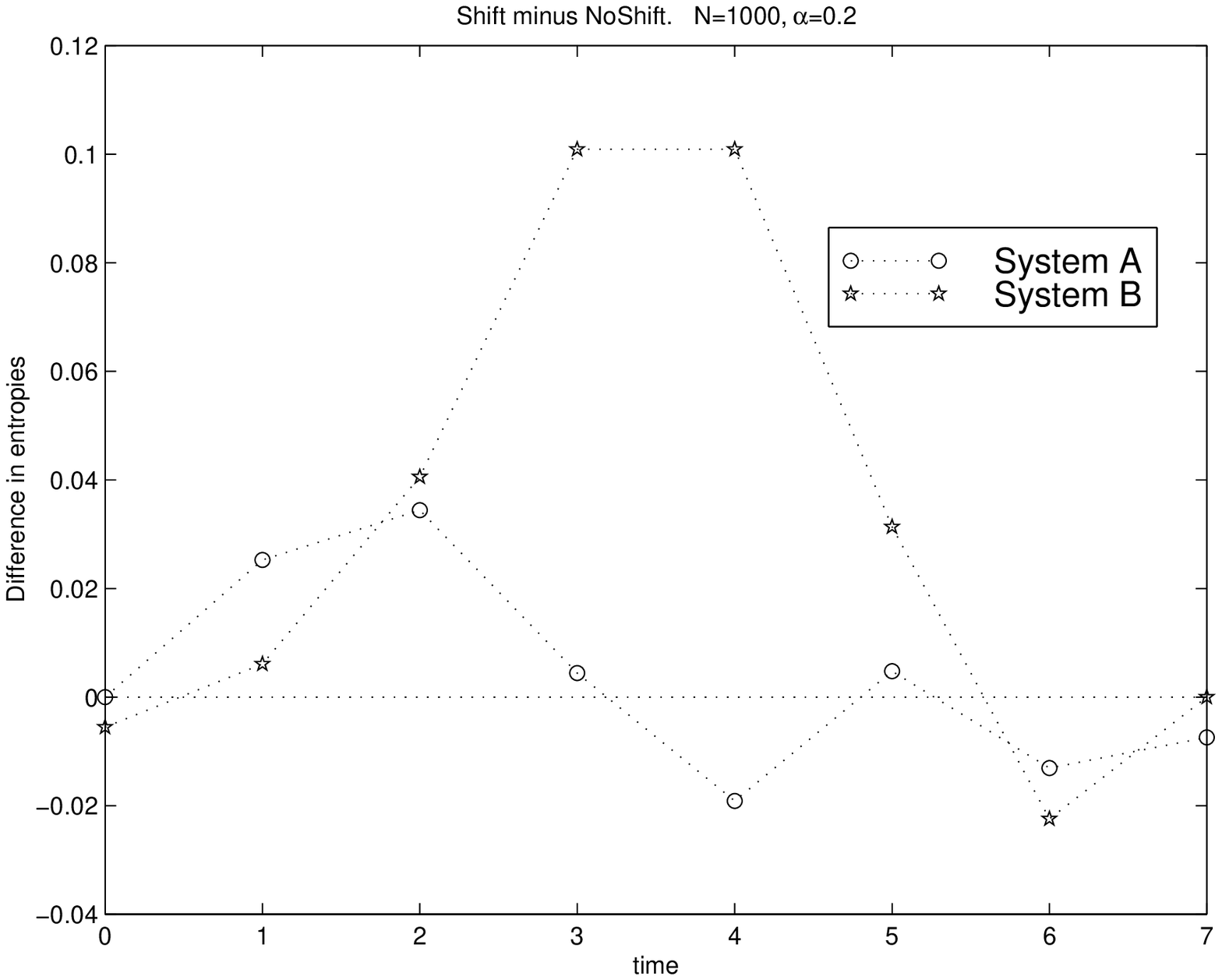}
}
\NI Figure 5. Difference in entropy between a run in which the evolution on time step \#4 is the identity versus one in which points of A (alone) undergo a shift (details in the text).
\vskip 1 pt
\hbox to 6 true in {\vrule  width 6 true in height .4pt depth 0pt}
}

First note that almost nothing happens to A. The shift is by integer multiples of 0.1 which is also the linear dimension of the coarse grains. But B does indeed respond to the ``wiggling" of A. However primitive this may be as a form of communication, this shows that wiggling A causes a macroscopic response in B.

\subsection{Possible physical origins of opposite-arrow regions}

\def\xxx{DTs(1-s)}

As discussed earlier, I believe the source of the thermodynamic arrow of time is the expansion of the universe \cite{gold}. Suggestions for the occurrence of opposite arrow regions will be developed in that same context. At the beginning of Sec.~4, I proposed one way to arrive at an opposite-arrow region that bucks the overall flow, namely a ``spaceship," or isolation. Thus you could have a region that has relatively little interaction with its surroundings, so that at a time closer to (say) the big bang, its most important conditioning would be from the big crunch end of the universe. There is a mathematical question hidden in this possibility. Specifically, if you condition a process to take a particular value at two widely separated times (0 and $T$), then in between it will tend to move away, and is likely to be farthest at $T/2$. Its entropy will be a symmetric function of time (cf.\ \cite{timebook}). By allowing spaceship survival and the like, I am introducing events for which a large excursion is moved away from the center. How unlikely is that? A much simplified way to examine this is to look at Brownian motion with diffusion constant $D$ that is conditioned on starting from 0 and returning to 0 at time-$T$. It is easy to see \cite{timebook} that the probability that it is at $x$ at time-$t$ is $1/[4\pi\xxx] \exp\left(-x^2/4\xxx\right)$, with $s=t/T$. Using this we can evaluate how unlikely various excursions are; for example, by integration it follows that $\Pr\left(x^2>\rho^2DT \hbox{~at~}t\right)= 1-\hbox{erf}\, \bigl(\rho/2\sqrt{s(1-s)}\bigr)$. If $\rho$ is not large this function imposes little relative penalty on deviations being displaced from $T/2$. For example, for $\rho$ as large as 5 (which is {\it extremely\/} unlikely---but {\it least unlikely\/} at $T/2$), it is only 3 times {\it more} unlikely at $0.4 T$. (At $T/2$ the probability is about $10^{-12}$.)

Starting therefore from the cosmo/thermo arrow connection, the only justification I can imagine for having opposite-arrow regions \cite{notantimatter} is living in a big bang-big crunch universe in which some regions manage the isolation needed to pass through the epoch of maximal expansion with arrow intact \cite{humility}.

\subsection{Detection of opposite-arrow regions}

What should this stuff look like, if it's there? By and large it should be extremely dull. Taking the opposite-arrow material to be ``relics" from the future, as discussed in Sec.~5.2, one must first consider how far off the big crunch is. Of course this is completely unknown, since there is at present no evidence that there will {\it be} a big crunch. (On the contrary, evidence today favors an ever expanding universe---but bear in mind the caveat in \cite{humility}). Just to grab a number, suppose the big bang-big crunch time separation is 500 Gyr. We are a mere 10--15 Gyr from the big bang and still see stars burning. But anything that's 485 Gyr old would be unlikely to have hydrogen burning and would probably be invisible at optical---or any other easily seen---frequencies. So this stuff would have gravitational interactions but be pretty much invisible, characteristics attributed to ``dark matter" \cite{darkmatter}. Like other dark matter candidates, this would be picked up in microlensing observations, which by themselves however give no indication of statistical properties.

Nevertheless, there are ways to make a positive detection. One possibility is to see an ``unexplosion." The idea is that even very old matter could have slowly developing processes that eventually lead to something dramatic---an explosion. From our perspective we would first see some fuzzy, relatively big region. It would gradually shrink in size, becoming brighter. Ultimately it would disappear.

A second method is to identify something that is {\it very\/} old. A way to achieve this is again to study the astrophysics of old objects, objects in which hydrogen burning has ceased, but in which other nuclear processes continue. Again these will be weak, for if they were strong they would long ``ago" have burnt themselves out (unless a threshold for something traumatic is achieved, as in the last paragraph). If specific nuclei can be identified for these processes, then there may exist particular gamma emission lines that characterize matter of such great age. 

A variant of this is the suggestion \cite{avignone} that one could extend a technique now used to date old rocks, namely isotope abundance. Thus if there is evidence that a particular isotope with a lifetime of say 10$^6$ year was present at a rock's creation, then its absence today can be taken as evidence that the rock is many times that age. Similarly, isotopes with lifetimes much greater are known (even $10^{21}$~yr), and finding appropriate samples depleted of these isotopes could be another indicator.

The search for observational evidence is what I consider the most significant challenge now connected with the newly-raised possibility of opposite-arrow regions of the cosmos.

\subsection{Cosmological implications}

The central problem in the modern study of cosmology is, is the universe open or closed? Will it continue to expand forever, or will it turn around and collapse? Finding an opposite-arrow region would be strong evidence for a big crunch. As indicated, it is difficult to think of any other source of opposite-arrow material. 

You could have a big crunch without a reversal of the thermodynamic arrow, and indeed early work on cosmology considered oscillating universes that warmed on each successive pass. Arguing from elegance, I consider that unlikely: if the geometry is symmetric and can nevertheless provide a thermodynamic arrow, why demand extra asymmetry ($\dots$ notwithstanding Pauli's remark: ``Elegance is for tailors").

As a small digression, I would like discuss other statistical ways to discern cosmological structure, supplementing material in \cite{timebook}. Consider a world with a big bang and big crunch, a world that must meet boundary conditions at both ends, boundary conditions that are far from equilibrium (or extremely unlikely) with respect to time periods interior to the time interval in question. To clarify, suppose the big bang and big crunch are at $\pm{\cal T}/2$ and we study an interval $\pm T/2$, with ${\cal T}> T$ and $({\cal T}-T)/2$ the duration of the interval from the big bang to the decoupling time. The homogeneous distribution of matter and its low metallicity at decoupling are natural when one works from an initial value close to the big bang. However, if one needs to meet this condition when {\it emerging\/} from a gravity-dominated universe and heading {\it toward\/} a big crunch, I expect it to be extremely {\it un}likely. Things must smooth out and you must get rid of all those big nuclei that formed along the way. So this is {\it not\/} an easy boundary condition to satisfy. One thing that would help would be to minimize the amount of processing of hydrogen, the amount that is caught up in burning stars. A state of the universe in which matter mostly did not accumulate in big masses, but stayed in bodies much smaller than the sun, would eliminate quite a bit of heavy element production, so that satisfying the boundary conditions we have described favors the existence of what would otherwise seem to be a disproportionate amount of nonluminous baryonic matter, for example, in the form of brown dwarfs.

Thus the prevalence of nonluminous baryonic matter, perhaps in the form of brown dwarfs, is consistent with the assumption of a geometrically and statistically symmetric universe. Could this ``consistency" be upgraded to evidence for a big crunch? Not quite, because at this point one really doesn't know what the abundance of such matter {\it ought\/} to be, reflecting for example uncertainties in just how many brown dwarfs should be created in an episode of galactic star formation.

\section{Paradoxes and causal loops}

In this section we take up the causal loops and paradoxes that could arise if opposite-arrow material is present and if communication between opposite running regions occurred. These paradoxes have been used \cite{penrose} to argue against Gold's idea, so it is worth confronting them here.

Consider the following tale of two observers with opposite-running thermodynamic arrows of time. We rename system A, Alice, and system B, Bob. The tale begins with Alice, who at 8 a.m.\ her time, sees rain coming in Bob's window (say this is 5 p.m.\ his time). With an hour's delay she informs Bob. He receives the signal at 4 p.m.\ his time and closes the window before the rain starts (say at 4:40p.m.\ his time). Did Alice see the rain coming in or not? 

The resolution of this paradox is similar to that in \cite{tachyon} and I have dealt with it at length in \cite{paradox}. The idea is that making the paradox precise requires setting up boundary conditions, in this case at two separate times, such that it would appear that no consistent intermediate motion could occur. Having done that, there are two possible resolutions. The first is that indeed your boundary conditions have no solution. From the perspective of the Setter of the boundary conditions this would mean that the world history He or She desired just doesn't happen. The second possible resolution uses assumptions of continuity and boundedness. In \cite{paradox} it is shown that in reasonable situations, including that suggested above in the carpet paradox, there will be some intermediate motion that {\it does} satisfy the boundary conditions.

For the soggy carpet paradox such resolution could take the following form. You set up Alice at 7:30 a.m.\ her time with instructions to send a message to Bob about his window, should that be necessary. You ``start" Bob at 3:30 p.m.\ his time with an open window. The self-consistent solution has Alice noting a {\it slightly\/} open window at 8 a.m.\ her time. Seeing this, Alice is unsure whether to send a message, and finally, at 9 a.m.\ her time, waffles, ``It will rain in a bit, you might think about closing your window." Bob gets this message and balances fresh air versus dry carpet by mostly closing the window, but nevertheless leaving it open a crack---which is just what Alice sees at 8 a.m.\ her time! This kind of self-consistent loop is favored by (what I consider) the better science fiction writers in their dealing with time-travel paradoxes \cite{heinlein}.

\section*{Acknowledgment}
This article is dedicated to Prof.\ Jean Vaillant on the occasion of his retirement. Much of this work was done at the Technion-Israel Institute of Technology. I am grateful to P. Facchi, E. Mihokova, A. Ori, S. Pascazio, A. Scardicchio, L. J. Schulman, and M. Roncadelli for helpful discussions. My research is supported in part by the United States National Science Foundation grant PHY 97 21459.


\begin{references}
\bibitem{kepler} ``Equilibrium" is not the right term here, since systems acting only under Keplerian forces do not have an equilibrium. All that is needed for our argument however, is that by becoming less homogeneous the system enters ever larger regions of phase space.
\bibitem{quotesense} The sense of the words ``since" and ``survive" in this sentence is with respect to the clock of the isolated region.
\bibitem{slice} In terms of our cosmological perspective, this time interval $[0,T]$ is far from both the big bang and the big crunch. The physical appearance of simultaneous arrows would be most likely to occur near the epoch of maximal expansion.
\bibitem{practice} The alternation of $\phi$ and $\psi$ enhances apparent time reversal symmetry. $\phi$ itself is not time symmetric, but for our purposes all that matters is that Lyapunov exponents for $\phi$ and $\phi^{-1}$ are equal (which is why Fig.~2b is symmetric). This is indicates \cite{timebook} that microscopic T violation alone need not provide a thermodynamic arrow.
\bibitem{notantimatter} An ``opposite-arrow region" consists of ordinary matter, differing from everyday matter only in statistical properties. It is not (necessarily) antimatter.
\bibitem{humility} Of course, when upwards of 90\% of the matter believed to be in the universe is in forms currently unknown to the human race, one must be humble in reaching conclusions from the absence of alternative explanations or in declaring the impossibility of a phenomenon. After all, the justification for the thermodynamic arrow that I find most appealing---expansion of the universe---could not have been guessed until that expansion was observed.
\end{references}
\end{document}